\documentclass[preprint,12pt]{elsarticle}

\usepackage{amssymb}
\usepackage{natbib}   
\usepackage{amssymb}
\usepackage{graphicx}
\usepackage{caption}
\usepackage{subcaption}
\usepackage{epsfig}
\usepackage{paralist}
\usepackage{hhline}
\usepackage{tabularx}

\journal{arXiv.org}

\begin{document}

\begin{frontmatter}



\title{A Distributed Multi-agent Market Place for HPC Compute Cycle Resource Trading}


\author[label1]{Stefan J. Zasada\corref{cor1}}
\author[label1]{Peter V. Coveney}
\cortext[cor1]{Corresponding author}
\address[label1]{Centre for Computational Science, Chemistry Department,
 University College of London,20 Gordon Street, 
London WC1H 0AJ, United Kingdom}

\begin{abstract}

Computer simulation is finding a role in an increasing number of scientific disciplines, concomitant with the rise in available computing power. Realizing this inevitably requires access to computational power beyond the desktop, making use of clusters, supercomputers, data repositories, networks and distributed aggregations of these resources. Accessing one such resource entails a number of usability and security problems; when multiple geographically distributed resources are involved, the difficulty is compounded. 

This presents the user with the problem of how to gain access to suitable resources to run their workloads as they need them. In this paper we present our solutions to this problem,   a resource trading platform that allows users to purchase access to resources within a distributed e-infrastructure. We present the implementation of this Resource Allocation Market Place as a distributed multi-agent system, and show how it provides a highly flexible, efficient tool to schedule workflows across high performance computing resources. 
\end{abstract}

\begin{keyword}

Grid Computing \sep Cloud Computing \sep Brokering \sep Multi-agent Systems \sep HPC



\end{keyword}

\end{frontmatter}


\section{Introduction}

Today's computational scientists face a growing number of challenges which affect their ability to fully exploit the computational resources available to them. Firstly, they have an unprecedented amount of computational power 
available to them, which will continue to grow in the future. A new 
generation of high performance computing (HPC) machines are now coming 
online with up to tens of petaflops performance. Secondly, the architectures of these large scale HPC machines point to a growing trend; HPC machines made up of hybrids of scalar and vector processors, or multicore processors that include scalar and vector components on the same chip, are likely to be commonplace in the future \cite{turek2007hpc}. This challenges application scientists to ensure their code is optimised to take full advantage of the hybrid architecture of a specific machine. The scale and complexity of high end HPC resources lead to users focusing on exploiting only a small subset of the resources available to them, and treating those resources as single islands of computational power. 

Distributed e-infrastructure \cite{cov} has sought to simplify end user access to and use of HPC resources, by establishing a common software platform for distributed computing conducted transparently across multiple administrative domains. However, the middleware tools developed to realize the computational `grid' concept have not always provided the transparency and ease of use envisaged \cite{chincov}. Essential to realising the vision of a distributed e-infrastructure as ubiquitous, seamless to use and as transparent as the electrical power grid, as proposed by Foster \textit{et al.} \cite{Anatomy}, is the broker. The broker or meta-scheduler is a component of a distributed e-infrastructure system responsible for efficiently distributing jobs between grid resources, taking into account factors such as machine load and cost models. A broker provides a point of contact between the user and the grid, placing application instances submitted by the user onto appropriate resources. The broker means that expensive HPC resources are used as efficiently as possible, ensuring that one machine is not idle while another has a large queue of jobs.

Subsequent to the development of grid computing has been the rise of cloud computing. Cloud computing adopts many different forms, but a unifying idea behind cloud computing is that a business model is used to monetize access to compute cycles in some way, and provide access to various resources such as CPU, memory and storage (known as Infrastructure as a Service clouds) and applications (Software as a Service or \textit{SaaS} clouds). For example, with so called Infrastructure as a Service (\textit{IaaS}) clouds, a user can gain access to a virtualized sever, and have complete control over that server as if it were his own machine, even though it is running in an administratively distinct domain. Cloud computing is a rapidly growing area due to major strategic investments from global software players such as Microsoft, Amazon, Google and IBM. Cloud storage today is thriving, particularly due to its shared data at low cost capabilities however there are many security and legal issues in cloud computing that are yet to be resolved. Typically, access to cloud resources is metered, and users must pay for the amount of CPU time or number of megabytes of storage that they use, with cloud computing users entrusting their data and software to third-party providers. Cloud providers may be commercial companies selling access for profit, or academic institutions, providing access under a research funding model. 

Within this paper, we use the term `distribute e-infrastructure' to mean any computing platform which implements some or all of the grid or cloud model; we believe that the software we describe herein is amenable to both. 

\section{Exploiting the Power of Distributed e-Infrastructure} 

In the high performance distributed e-infrastructure domain, all too often the user's time is spent investigating the availability of resources, marshalling data and minding their applications. For most computational scientists using high performance computing, distributed e-infrastructures have failed to deliver their promise of providing transparent, ubiquitous computational power on demand. This is due to a lack of both appropriate tools and ones that present the right level of abstraction to the user \cite{chincov}, meaning that it is easier for users to carry on with their existing usage patterns, as if the e-infrastructure was not there. When the user has access to more than one computational e-infrastructure, each running a different middleware stack, the problem is compounded, with the user having to learn how to use different middleware client tools to interact with the multitude of resources available to them. 

HPC resource providers have been chasing ever increasing machine peak performance, with several petaflops machines now commonplace. However, the end user, the so called `application scientist', is less interested in the peak performance of the machine they are using; their primary concern is the total wallclock time to solution of the scientific problem that they are working on \cite{PromitaChakraborty06282009}. Many strategies have been employed to try to reduce the total time to solution, although they are of course highly dependent on the nature of the problem being solved. A second concern of the user of a distributed e-infrastructure platform is the cost of running their applications, the problem of choosing a resource (or multiple resources) in order to perform a simulation becoming a trade-off between the total cost of running the simulation and the wallclock time to achieve a result. 

\section{Addressing Users' Problems}

We believe that with the right combination of tools and services, the original concept of a distributed e-infrastructure as a provider of transparent and ubiquitous computing power can be realised. To this end, we have developed the Application Hosting Environment \cite{zasada2014flexible, aheusability}, a lightweight interface layer that provides a higher level of abstraction than many other middleware tools, to allow the user to concentrate on running their applications without having to worry about the minutiae of dealing with every possible combination of compiler, architecture and queuing system. 

However, AHE still requires users to choose the individual machines on which to run, and does not provide capabilities to help users to minimise their time to solution. We have therefore developed a flexible decentralised workload allocation system that implements a controlled computational market place, to enable the trading of time on HPC resources and allowing the user to control the aspects of the workload that they are interested in: the cost and the time to solution.

We believe that this decentralised system is more scalable than currently available resource brokering technologies and is able to more efficiently allocate work between a set of resources based on cost minimization and run time optimisation. The decentralised nature allows resources to easily join and leave the system, potentially creating dynamic virtual organisations based on aggregated resources from federated resource providers. Our system is based on a combinatorial, multi-attribute reverse auction mechanism, which we describe below. 

\section{Market Place Requirements}
\label{mechdes}

We performed a user needs analysis by examining common resource usage patterns to provide a basis for the design of our resource allocation market place. Key features relating to the resource allocation mechanism are that it must be user initiated, capable of allowing users to specify their requirements for an application run, and permitting users to request access to multiple resources, specifically: 

\begin{enumerate}
 \item The process of submitting an application should be initiated by the user and done at the user's convenience, rather than at a time specified by the computational resource provider. 
 \item With current systems the onus is on the user to choose the resource on which they want to run, meaning that they often choose the one they think will be able to run their job fastest, or the one they are most comfortable using. Instead of requiring users to choose resources, our system allows the user to specify requirements for their application to run, such as the time they need the results to be produced by, or the maximum cost they are willing to pay in order to run the application. 
 \item Users may require access to multiple resources in order to run their application. For example, for an application that consists of a simulation code and a coupled visualization engine, the user would need access to a compute resource and a visualization resource \cite{zasada2012distributed}. 
\end{enumerate}

Arising from these requirements we have developed a resource allocation system based on a combinatorial, multi-attribute reverse auction, in which resource providers compete for workloads offered by users. We name this system the Resource Allocation Market Place (RAMP)\footnote{The RAMP system software has not yet been publicly released, but is available from the authors on request.}.  

\subsection{Design Constraints}

In order to satisfy the requirements discussed above while maintaining a focus on usability, we have placed the following constraints on our resource allocation system:

\begin{itemize}
\item The user should not have to configure the details of every resource that he may wish to use. The system should automatically discover suitable resources as they become available. 
\item The user should be able to specify what he requires from a resource in order to run an application. Any requirements explicitally specified must be met by the responding resource. However, if no resource can satisfy the requirements after \textit{N} rounds of bidding, a resource can make its best offer to the user. 
\item The RAMP system is accessible through the same client used to access the AHE. The user should not be concerned with any details of where or how an application is run. 
\end{itemize}

\section{Developing the System}


Multi-agent systems (MAS) and distributed e-infrastructure environments possess a number of complementary features \cite{C1}. In addition to the agent programming paradigm, agent development environments such as JADE \cite{39} provide a framework in which much of the software tooling required to develop agents, establish inter-agent communication and so on is already provided, much simplifying the process of developing multi-agent systems. 

In the context of the reverse auction based metascheduler, multi-agent systems also bring a number of other benefits. One of the drawbacks of some of the meta-schedulers is that they rely on central information services to aggregate data from resources and maintain their world view. The obvious limitation of this approach is that scheduling decisions may be made on out of date data. In the MAS approach proposed, each agent is responsible for maintaining the view over its own sphere of the world, meaning that the data used to make scheduling decisions is more current. This accords with the devolved nature of a distributed e-infrastructure, and especially federated e-infrastructures.

The application of multi-agent systems to auctions discussed above shows that trading systems and economies can be successfully built from interacting software agents. This model is thus also applicable to the distributed e-infrastructure economy that is developing as commercial providers trade compute power on an open market (badged as `cloud' HPC). Finally, MAS provides a software development framework featuring a high level of abstraction for building autonomous, rationally functioning software systems. The distributed nature of e-infrastructure systems coincides with the distributed, multi-agent system model of programming, and leads to the development of fault tolerant peer-to-peer systems, in which the failure of one components does not have a fatal impact on the rest of the system. 

This makes the MAS paradigm ideally suited to develop our distributed resource allocation system. Within the system, software agents can act on behalf of the different entities involved, principally users and resources. Two different BDI (i.e. one characterized by its beliefs, desires and intentions) types of agents feature in the system:

\begin{itemize}
\item Resource management agents, responsible for maximising the utilization of a resource. A Resource Agent is run on each constituent resource of the distributed e-infrastructure. It maintains a predictive model of resource availability, which it uses to decide when it is able to run a job. It can vary the cost of the offers to run jobs that it makes to encourage jobs to run when the machine is free by lowering its prices, or increasing the cost when the machine is overloaded with jobs to maximize revenue, based on a set price range configured by the resource administrators. 
\item User Agents, responsible for gaining access to resources at a cost and availability specified by the user. The User Agent runs on the client machine, and negotiates with the resource management agents for the most appropriate resource to run a particular application. Users provide the agent with a description of their cost and time requirements for the job; they can either ask for the job to be run in the fastest time possible, at the least cost, or at a specified maximum cost and/or wait time. The agent's goal states consist of minimising the cost of the job, minimising the wait time of the job, or at least matching the specified requirements. The agent then initiates multiple rounds of bidding with the resource management agents until the requirements are achieved and the application is launched, or if the requirements cannot be met the user is presented with the best offer received. If no offer is received, the application fails to run. The user will be able to specify both static and dynamic constraints on their job, as defined in a Request for Quotation Language (RFQL) schema which we have developed. 
\end{itemize}

In addition, a Banking Agent acts as a collation point for all successfully actioned requests. 

\subsection{Developing the Negotiation Protocols}
\label{sec:protocol}

The process of participating in a reverse auction requires the agents involved to communicate in a structured way. Fortunately, a standardized way exists to achieve this. The Foundation for Intelligent, Physical Agents (\textbf{FIPA}) exists to develop standards relating to software agent technologies. The standards that FIPA develop provide a mechanism for software agents to be mutually understood, regardless of underlying implementation technologies. The FIPA Agent Communication Language (\textbf{ACL}) \cite[pages-10-17]{jadebook} specifies the structure of inter-agent messages, and defines a set of \textit{communicative acts} (CAs), performed by the act of communicating. These CAs, along with a bespoke content language, allow agents to participate in a reverse auction. 

We define three different procedures for agents to communicate in different circumstances:

\begin{itemize}
\item The reverse auction negotiation - the actual negotiation process required to conduct a reverse auction.  
\item The banking update negotiation - the process of notifying the Banking Agent to record a successful auction result. 
\item The cancellation negotiation - the process of a user cancelling a request. 
\end{itemize}


\subsubsection{The Reverse Auction Protocol}
\label{sec:algorithm}
The reverse auction algorithm developed is adapted from that described by Matsuo \textit{et al.} \cite{matsuo2002dbr}. The auction consists on \textit{N} rounds of open bidding, where all sellers can see the bids made by other sellers. As the auction must be based on multiple attributes, the user is able to specify their requirements through a request for quotation (described in \S \ref{sec:rfql}). 

The auction is combinatorial, meaning that multiple units can be requested. Each sub-request should be treated as a separate auction in the system. This means that an inconsistent state could result, where some parts of an overall request are successful and others are not. Therefore, the auction protocol incorporates a two-phase commit process to ensure the availability of all requested resources. 

\begin{figure}[!t]
\centering
\includegraphics[scale=0.4]{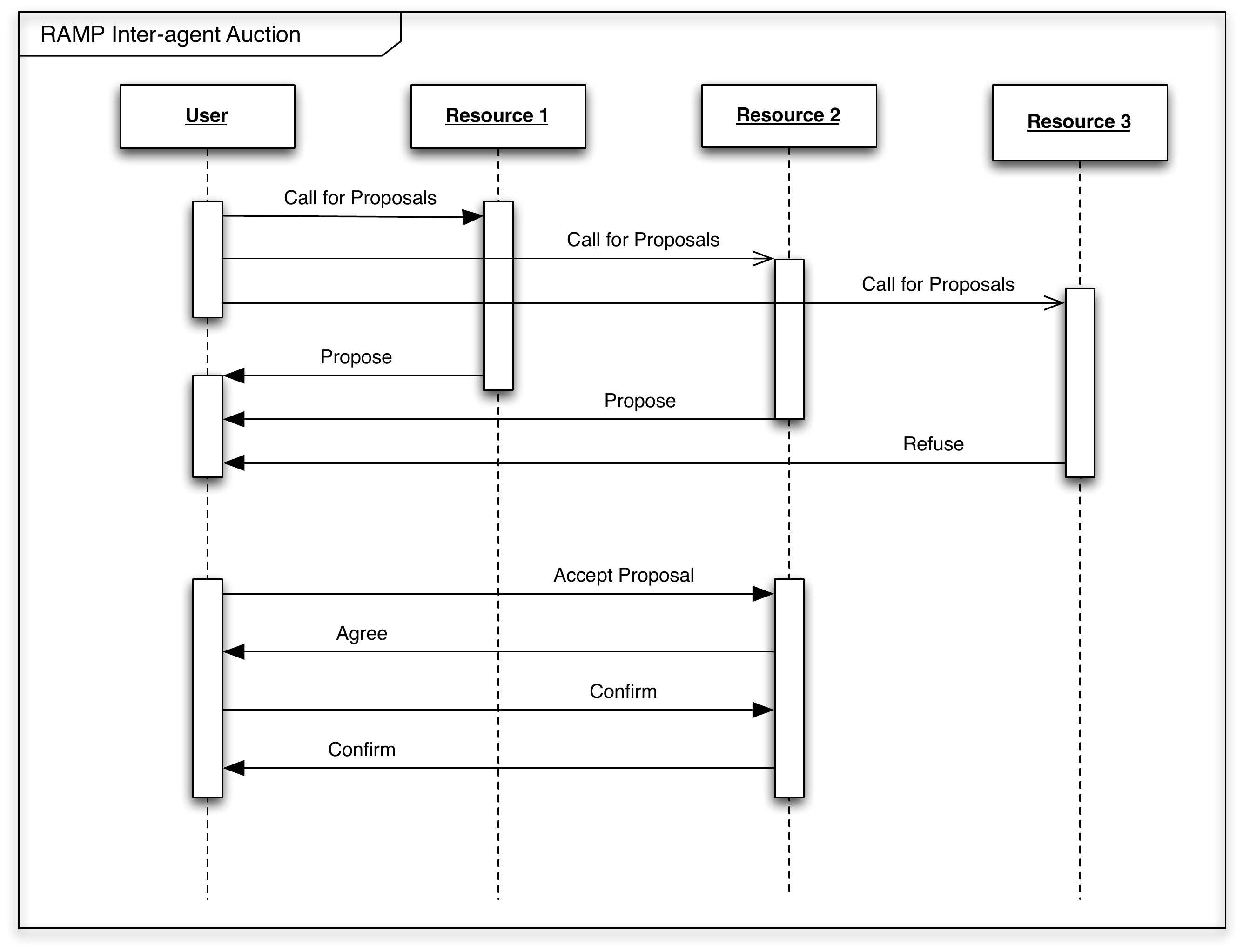}
\caption{The sequence of messages between a User Agent and three resources required to implement a single round of bidding for a single unit auction. Inter-agent communication is performed using messages specified in the FIPA protocol. In this example, Resource Agents 1 and 2 respond to the RFQ with offers to accept the workload, while agent 3 refuses. The User Agent accepts the offer from agent 2 using a two phase commit.}
\label{fig:auction}
\end{figure}

Briefly the algorithm flow is shown in figure \ref{fig:auction}, and is as follows:
\begin{enumerate}
 \item A User Agent initiates the auction by advertising their requirements with Resource Agents via an RFQ (\texttt{FIPA: Call for Proposals}).
\item Resource Agents evaluate the RFQ and decide whether they can satisfy the request (or section of a request in the case of a combinatorial request), based on their utilization and CPU hour cost. The Resource Agents which can satisfy the request make bids, which are propagated to the User Agent (\texttt{FIPA: Propose}). If the Resource Agent cannot accept the request, it notifies the User Agent (\texttt{FIPA: Refuse})
\item The User Agent evaluates the requests it has received, and stores them in a ranked list if they meet its requirements, or else rejects them to the submitting Resource Agent (\texttt{FIPA: Reject Proposal}). When the next round of bidding commences, the User Agent modifies its RFQ to correspond to the best offer it has so far received, which is sent to the Resource Agents as its revised request. Steps 2 and 3 are repeated \texttt{N} times. 
\item After \texttt{N} rounds of bidding, the User Agent evaluates the final set of bids received.
\item The User Agent selects the most optimal bid or bids that match requirements and notifies the winning Resource Agents(s) (\texttt{FIPA: Accept Proposal}) or, if no bid or bids match the requirements, the closest matching set are presented to the user for approval. The user can also configure the system to allow manual approval of all bids.
\item \label{step:agree} The Resource Agent(s) holds a slot for the winning bid on the machine queue by creating a reservation in the queuing system, and sends an acknowledgement back to the User Agent that they are willing to proceed (\texttt{FIPA: Agree}) along with a reservation ID for the requested slot on the relevant resource. If the Resource Agent cannot now satisfy the request (because more jobs have been queued on the system in the meantime for example), the agent withdraws from the auction (\texttt{FIPA: Refuse}). This is the \textit{voting phase} of the two-phase commit. 
\item The User Agent works through all offers received until all parts of the request have been agreed to. Where a winning bid is subsequently refused, the User Agent contacts the next best bid and so on until all available bids are exhausted. 
\item If all sections of a request are agreed to, the User Agent notifies all successful Resource Agents (\texttt{FIPA: Confirm}) and sends a digitally signed copy of the RFQ and the reservation ID back to the Resource Agent, signed using the user's personal X.509 credential. This establishes that the user has agreed to the reservation.  The Resource Agent acknowledges this message (\texttt{FIPA: Confirm}). This is the \texttt{commit phase} of the two-phase commit. 
\item If all sections of a request cannot be agreed to, the User Agent cancels all requests received (\texttt{FIPA: Cancel}).
\item When a Resource Agent receives the (\texttt{FIPA: Agree}) message (step \ref{step:agree}) and creates a reservation slot in its queuing system, it begins a timer process. If the Resource Agent does not go on to receive a (\texttt{FIPA: Confirm}) message within a given time period, it cancels the reservation slot in the queue. 
\end{enumerate}

\begin{figure}[!t]
\centering
\includegraphics[scale=0.4]{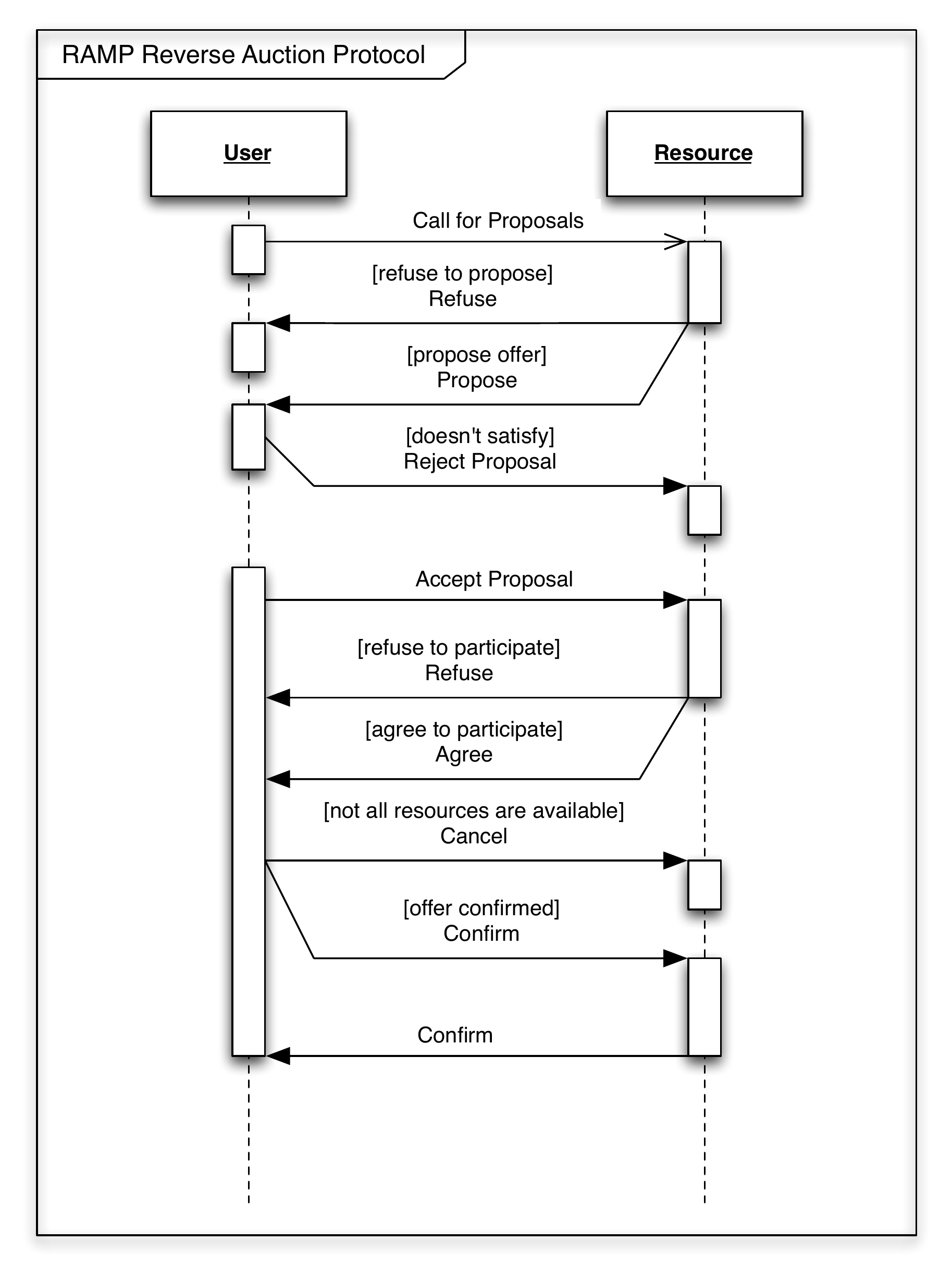}
\caption{The sequence of FIPA messages between a user and a resource in an auction negotiation with \textit{n=1} rounds of bidding.}
\label{fig:reverse}
\end{figure}

This is the protocol employed by the user and Resource Agents to negotiate access to computational resources at user specified time periods. The sequence of FIPA operations are shown in figure \ref{fig:reverse}.

\subsubsection{Reservation Notification Protocol}
\label{sec:notification}

This protocol is used to inform the Banking Agent that a request has been successfully fulfilled. After a successful negotiation, the Resource Agent proceeds as follows:

\begin{enumerate}
\item The Resource Agent takes the signed message from the User Agent, containing the RFQ and reservation ID, and digitally signs it itself, using its own certificate. 
\item The Resource Agent sends this signed document to the Banking Agent (\texttt{FIPA: Request}).
\item The Banking Agent confirms the digital signatures applied to the message to establish that the veracity of the message, and then debits the user's account in accordance with the request and credits the resource's account commensurately. It then notifies the requesting Resource Agent (\texttt{FIPA: Agree}).
\item If the Banking Agent cannot validate either signature, it responds to the requesting Resource Agent with (\texttt{FIPA: Refuse}).
\end{enumerate}

\begin{figure}
\centering
\includegraphics[scale=0.4]{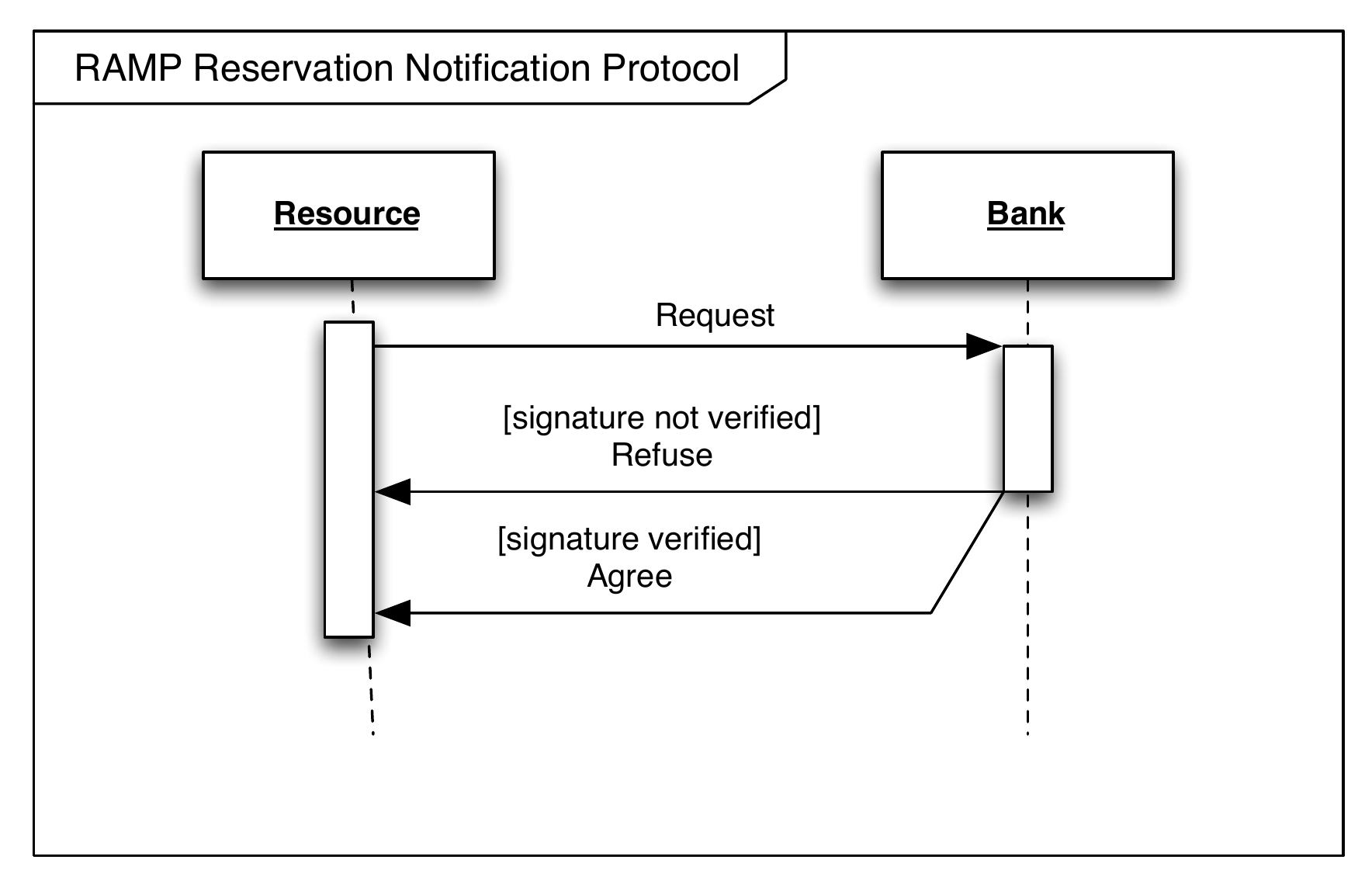}
\caption{The sequence of FIPA messages between a resource and a Banking Agent required to notify the bank of a successful negotiation.}
\label{fig:notify}
\end{figure}

The sequence of operations is shown in figure \ref{fig:notify}. If a user needs to cancel a reservation once made, they can do so through the RAMP system. If the Resource agrees to the cancellation, the user's account will be re-credited with the cost of the resource slot.

\subsection{Specifying a Quotation Request}
\label{sec:rfql}

As we discussed earlier, the overriding concern of the HPC e-infrastructure user is the time to solution for the problem that she is working on, with a further concern of how much the application will cost to run. Our task is to identify the terms which a user needs to specify her requirements from a machine in order to run her application, and to capture these terms in a request for quotation language (RFQL) which provides a standard way of requesting quotations to run applications from resources. Since the AHE takes care of maintaining information such as which resources have which applications installed, the RFQL need not contain terms to specific to the instantiation of an application the user wants to run (such as the location of a binary); instead it need only contain the terms required to describe requirements from a machine to run the application. 

Our language also needs to contain terms to allow the user to specify cost and deadline requirements, and aspects that might affect the performance of the application, such as operating system running on the resource, the maximum RAM available, or the CPU (or GPU) architecture. The terms used in our RFQ language are described below:
   
 \begin{itemize}
 \item \textbf{CPUHourCost} - the maximum cost per core hour that the user is prepared to pay in order to run her application. 
 \item \textbf{EndDate} - the date by which the user requires her application run to be complete. 
 \item \textbf{EndTime} - the time by which the user requires her application run to be complete.
 \item \textbf{StartDate} - the time after which the user needs her application (or workflow component) run to start. This is useful if the application run is part of a workflow and depends on a previous application run completing before it is able to start. 
 \item \textbf{StartTime} - the time at which the user needs her application to start. 
 \item \textbf{OperatingSystem} - the operating system that the user requires the grid resource to be running. 
 \item \textbf{OSVersion} - the version of the operating system that the user requires the execution resource to be running. 
 \item \textbf{Architecture} - the CPU/GPU architecture that the user requires the execution resource to consist of. 
 \item \textbf{CPUSpeed} - the minimum CPU/GPU speed that the user requires of her target resource. 
 \item \textbf{WallTime} - the maximum time that the application will run for. 
 \item \textbf{TotalDiskSpace} - the total disk space that the user needs to be available in order to run her application (this term and the NodeDiskSpace term are mutually exclusive).
  \item \textbf{NodeDiskSpace} - the total disk space available on each compute node (this term and the TotalDiskSpace term are mutually exclusive).
 \item \textbf{InterNodeBandwidth} - the minimum network bandwidth that the user requires between the nodes on the target resource.
 \item \textbf{RAMPerCore} - the minimum amount of RAM that the user requires to be available per compute core. 
  \item \textbf{TotalCores} - the total number of compute cores that the user needs to have access to (this term and the NodeCount/NodeCores terms are mutually exclusive).
 \item \textbf{NodeCount} - the number of compute nodes that the user requires access to (this term and the TotalCores term are mutually exclusive).
 \item \textbf{NodeCores} - the number of cores per node that the user requires access to (this term and the TotalCores term are mutually exclusive).
 \end{itemize}
 
 These terms are expressed using XML syntax, formally defined by an XML schema. Several of the attributes are required in each instance of RFQL: CPUHourCost, EndDate, EndTime and either TotalCores or both NodeCount and NodeCores. The other attributes are optional, and it is assumed that the user is not interested in making a decision based on any attribute which is not specified. If an attribute is present, a resource must be able to satisfy it before responding to the RFQ. 

The schema allows for multiple requests to be made within a single RFQL document, meaning that a user can request a combination of resources in order to perform a workflow, or run a highly distributed application. No formal mechanism is provided to specify dependencies between individual requests, but the StartTime and StartDate terms allow the user to request resources sequentially in time. In this way, the user is able to specify complicated advanced (co-)reservations for time on one or more resources. 
 
The approach we have taken with RFQL is to define a small vocabulary that captures the computational requirements that the user is interested in. This compares with the Condor ClassAds approach, which allows users and resource providers to define arbitrary terms  in the job descriptions. We believe that our small, well defined vocabulary is the correct approach to take here, as it aids system development and improves the likelihood of request/resource matching. We do not expect users to code RFQL by hand, but instead generate documents automatically through our interface tooling. 

\subsubsection{Dynamic Service Level Agreements}
 
An RFQ, plus a response from a resource that satisfies the request, constitutes a contract between the user and the resource to allow the user access to the specified number of processor cores on the resource, for the specified period of time. It can be considered a dynamic service level agreement (SLA) to provide a specific, one time service to a user at a defined cost. Enforcement of the SLA is beyond the scope of this paper however. 
 
\section{Implementation}
 
We used the agent specifications and communication protocols, along with the Request for Quotation notation, to implement a multi-agent system in Java using the JADE framework. JADE was chosen because it provides a distributed agents platform, supports coordination between different FIPA compliant agents and provides a standard implementation of the FIPA-ACL communication languages. Agents can be quickly constructed by extending the \texttt{jade.core.Agent} class. The capabilities of the agent are then defined by implementing behaviour classes which extend subclasses of the \texttt{jade.core.behaviours.Behaviour}.

Below we review the three agent types we have defined, and discuss the implemented behaviours that provide their capabilities. 

\subsection{User Agent}

\begin{figure*}[!t]
\centering
\includegraphics[scale=0.3]{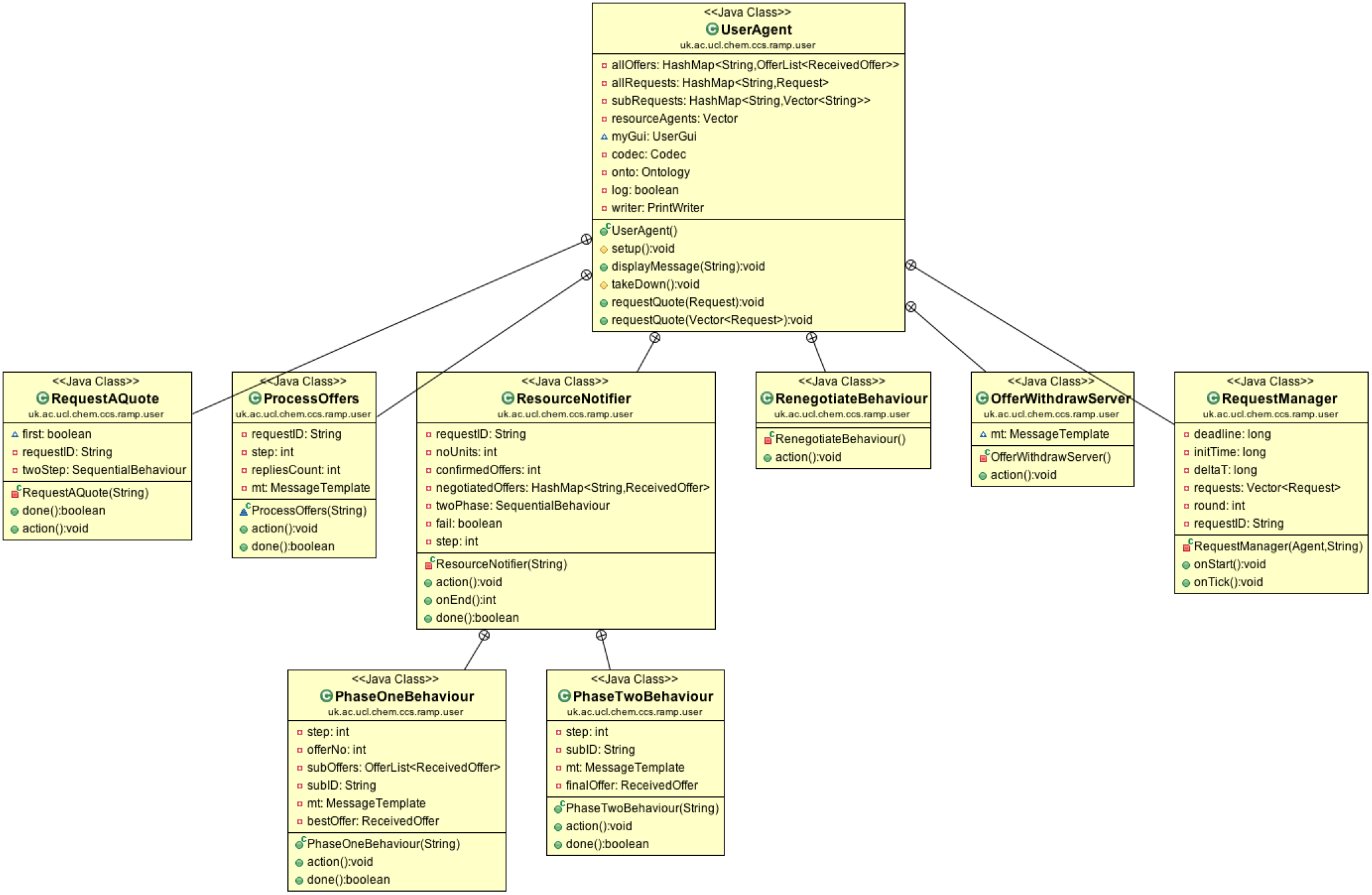}
\caption{The hierarchy of behaviours implemented by the User Agent.}
\label{fig:useragent}
\end{figure*}

As mentioned, the User Agent is responsible for purchasing resources on the instruction of its owner. Each user has a single User Agent to manage their requests. Since it initiates and manages the auction process, it is the most complicated agent, comprising the greatest number of behaviours. The hierarchy of behaviours is shown by the class diagram presented in figure \ref{fig:useragent}.

\begin{itemize}
\item \texttt{RequestAQuote}: This behaviour is responsible for taking user requests in the form of RFQ documents, translating them to the inter-agent communication ontology used by the RAMP system and imitating and managing the rounds of bidding in the auction. The number of rounds and intervals between rounds are user configurable parameters. 
\item \texttt{RequestManager}: The behaviour is used by the \texttt{RequestAQuote} behaviour to manage the individual rounds of bidding. It extends \texttt{jade.co\-re.behaviours.TickerBehaviour} to trigger an event (another bidding round) at set time intervals. 
\item \texttt{ProcessOffers}: This behaviour processes offers received from Resource Agents during the bidding process, and is responsible for sorting the offers received. A simple sorting algorithm is used, sorting first on cost, then deadline, then the order offers are received in. 
\item \texttt{ResourceNotifier}: This behaviour executes once the \texttt{RequestAQuote} behaviour completes, and is responsible for finalising the purchase of the resources requested. It implements the two phase commit required to ensure consistency when purchasing multiple units in an auction through two sub behaviours:
\begin{itemize}
\item \texttt{PhaseOneBehaviour}: This behaviour implements the \textit{voting phase} of the two phase commit, instructing successful resources that the user would like to accept their offer, and processing confirmations or rejections from those resources. 
\item \texttt{PhaseTwoBehaviour}: This behaviour implements the \textit{commit phase} of the two phase commit, confirming the offer acceptance if all required resources are available, or cancelling the transaction if not. 
\end{itemize}
\item \texttt{RenegotiateBehaviour}: This behaviour is used by the User Agent to initiate the cancellation procedure.
\end{itemize}

The user initiates resource auctions by passing the User Agent one or more RFQ documents. To simplify this process, the User Agent provides a graphical user interface, shown in figure \ref{fig:rampgui}. This GUI allows the user to load RFQ documents, initiate and monitor auctions, and also view and manage purchased resources. 


\subsection{Resource Agent}

\begin{figure}[!t]
\centering
\includegraphics[scale=0.45]{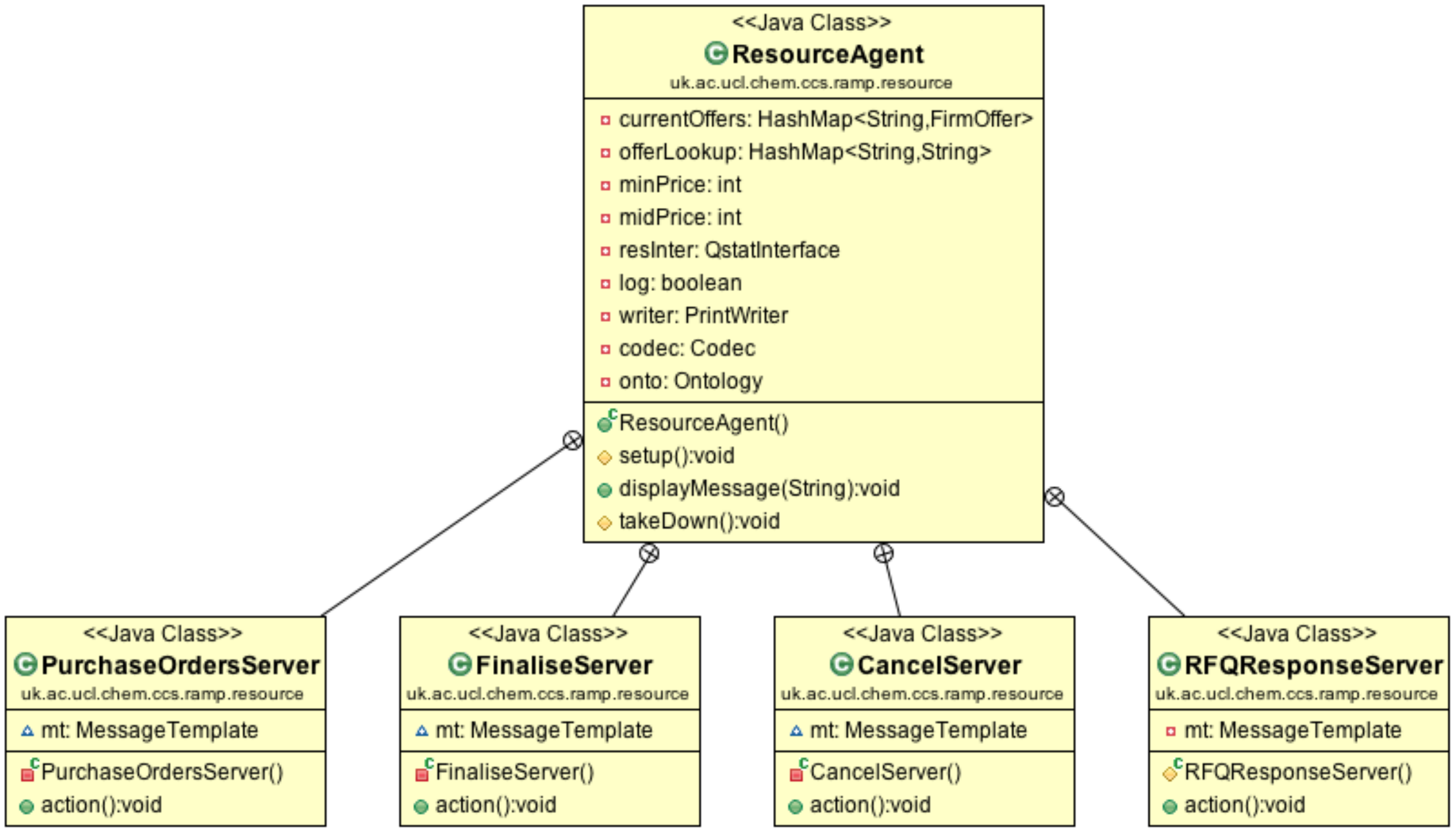}
\caption{The hierarchy of behaviours implemented by the Resource Agent.}
\label{fig:resourceagent}
\end{figure}

Each resource that participates in the resource allocation market place runs a resource management agent, which is responsible for responding to requests for quotation and negotiating the sale of CPU/GPU time. In order to do this, the Resource Agent implements four distinct behaviours, shown in figure \ref{fig:resourceagent} and described below:

\begin{figure}[!t]
\centering
\includegraphics[width=1\linewidth]{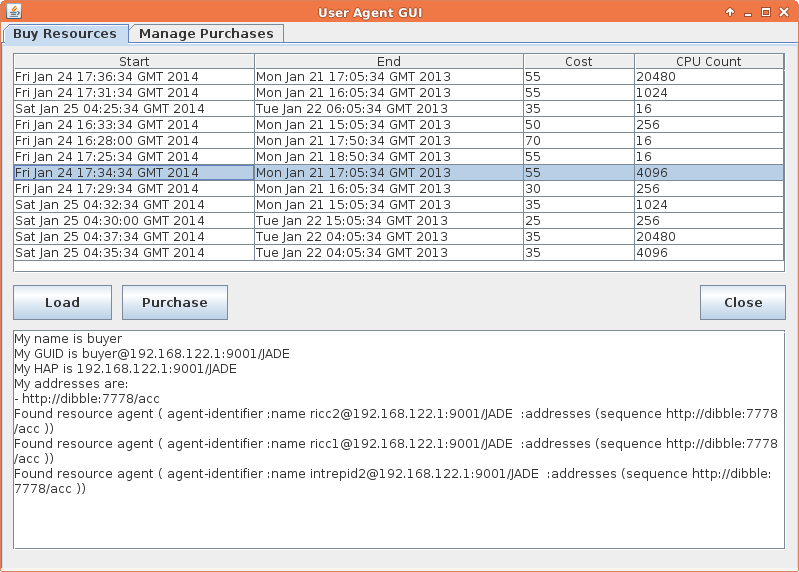}
\caption{The User Agent graphical user interface.}
\label{fig:rampgui}
\end{figure}

\begin{itemize}
\item \texttt{RFQResponseServer}: This behaviour listens for requests for quotations made by User Agents, evaluates those requests and then either submits an offer in response to the request, or else declines to participate.  
\item \texttt{PurchaseOrdersServer}: This behaviour listens for the acceptance of offers from User Agents. When an offer is accepted, this behaviour creates a tentative reservation within the machine's queuing system to correspond to the offer. If the requested resource is no longer available, this behaviour declines the acceptance of the offer. It implements the voting phase of the two phase commit protocol on the Resource Agent side. 
\item \texttt{FinaliseServer}: This behaviour listens for messages from the User Agent confirming the finalization of an offer. When an offer is finalized, it stops the time out set on the queue reservation, locking it in. It implements the commit phase of the two phase commit protocol on the Resource Agent side. Once an offer is fully confirmed, this behaviour is responsible for initiating the Reservation Notification Protocol described in Section \ref{sec:notification}.
\item \texttt{CancelServer}: This behaviour provides the cancellation capabilities.
\end{itemize}

\subsubsection{Modelling system availability}

The Resource Agent maintains an internal representation of the resource that it manages in order to be able to respond to request for quotation. Obviously, the resource must know details of the resource in terms of the CPU/GPU types available, memory per node and so on. The administrator of the resource therefore configures these static properties via a configuration file, prior to running the Resource Agent. These static properties correspond to the terms of the RFQ specification (\textit{cf.} \S \ref{sec:rfql}), but exclude the two dynamic properties relating to price and time, which the Resource Agent derives from the queuing system.  

Note, we consider a resource to be made up of homogeneous compute nodes, although this does not always conform to reality, whereby a large HPC system could be made up of CPUs/GPUs/nodes with different speeds, nodes with different memory sizes and so on. Systems that comprise heterogeneous architectures can be supported by running multiple instances of the Resource Agent, one for each distinct part of the machine. 

The Resource Agent will examine each RFQ that it receives, and then decide whether to make an offer or not. Evaluating a request involves two steps:

\begin{enumerate}
\item First, the Resource Agent checks the static terms of the request (such as the requested CPU type) against its internal resource model. If the resource cannot satisfy the this part of the request, then the Resource declines the offer to participate. 
\item Next, the Resource Agent examines the current load on the resource. If sufficient free CPUs exist at the point in time that they are required, the Resource Agent calculates an offer price (see below), checks that this offer price meets the request, and then makes an offer. 
\end{enumerate}

\subsubsection{Interfacing with the Queuing Systems}

The Resource Agent must interact with the resource it manages to obtain a view of system utilization which can be used to respond to requests for quotation, and generate offer prices for those responses. The default implementation interfaces directly with the queuing system to obtain a measure of the load on the resource (in terms of available CPUs) at the point in time when the requested job must be satisfied. While providing an adequate model of system usage, the default queuing system is not without its drawbacks, in that when examining the queuing system to evaluate future availability, it makes calculations based on the wall time specified by the user for each running job. Often, a user will use the system default wall-time, meaning that a job could finish long before the queuing system expects it to. Obviously, the queuing system copes with this by running the next job in the queue, but this can lead to situations where the Resource Agent expects the system to be unavailable when it is not. 

However, the Resource Agent features a plug-in architecture that allows the resource administrators to substitute an alternative resource interface where available. For example, a resource administrator could implement an interface that queries a resource availability modelling service, such as the QBETS batch queue prediction service \cite{nurmi2008qbets}. The QBETS service can be used to predict a statistical upper bound on how long the job is likely to spend waiting in the queue prior to execution and given the job characteristics and a start deadline, can calculate the probability that the job begins execution by the deadline. The Resource Agent can then use this information to make offers, rather than relying on the basic queuing system interface. 

\subsubsection{Pricing model}
\label{sec:primod}

The Resource Agent is responsible for setting the offer price made to resource requests coming from User Agents. The price offered is varied based on the load on the machine at the time the job must be run. If the machine is lightly loaded, the resource makes a low offer, in order to attract more work to the machine. If the resource is heavily loaded, the machine offers a higher price, or if it is saturated, refuses to participate in the auction at all. 

These prices depend on two resource administrator defined parameters, the \texttt{start price} and the \texttt{minimum price}. As its name suggests, the minimum price is the lowest price a Resource Agent will ever offer in an auction. These values are used to calculate the decrement by which the request is reduced by the resource when making an offer, according to the following formula

\[ {dec} = {(sp-mp) \over s \times (1-l)} \]

\noindent where \(sp\) is the starting price, \(mp\) is the minimum price, and \(l\) is the percentage of the machine that will be allocated at the time the job is to be run, expressed as a decimal. Although the number of bidding rounds is controlled by the User Agent, the Resource Agent anticipates that there will be multiple bidding rounds using the \(s\) parameter, so that the total decrement is not applied in one go, but gradually over several bidding rounds. This gives a value to decrease the request price by, \(dec\), which results in a bespoke spot price for the resource at a given point in time and in response to a user's request. 

\subsection{Banking Agent}

The Banking Agent is tasked with recording transactions between users and resources, and can provide an overview of the overall system of resource trading. To do this it implements three behaviours, described below:

\begin{itemize}
\item \texttt{TransactionUpdateBehaviour}: This behaviour listens for and processes update messages from the Resource Agents on the completion of successful transactions, to update the internal balances of Resource and User Agent. 
\item \texttt{CancelListenerBehaviour}: This behaviour listens for and processes cancellation messages, and notifies both User and Resource Agent when the cancellation is complete. 
\item \texttt{BalanceRequestBehaviour}: This simple behaviour can provide a user or resource with a statement of their balance on receipt of a digitally signed request.
\end{itemize}


In order to verify the digital signatures appended to transaction update messages, the Banking Agent maintains a record of the public key of participants in the market place. This is done when accounts are credited be the Banking Agent administrator. Within the RAMP system, a standard virtual currency is used. 

\subsubsection{A Note on Banking}

The Banking Agent is designed and implemented as a technical way to keep track of deals made between users and resources, but it is not intended to answer the policy questions that address how usage of resources is reconciled with real-world cash payments, which is beyond the scope of this paper. Some of those policy questions relate to how the currency used in the system converts to real world currencies, how deals are enforced, and how overdrafts can be dealt with, and provide a rich and interesting vein of future work. Within a production system, we envisage that one or more Banking Agents will be run by an independent, trusted third-party. 


\subsection{Inter-agent Communication}

Peer to peer (P2P) systems are recognized as a way of building large, scalable distributed systems. Like distributed e-infrastructure systems, they have evolved as a way to share resources across administrative domains, but do so from a very different starting point and with very different requirements, in terms of security and availability \cite{trunfio2007peer}. A key feature of the RAMP system compared to other resource brokering/meta-scheduling systems is that there is no central service in overall control of the system. In effect, the Resource Agent and the User Agent are peers in a peer to peer system, and connected together by a P2P network infrastructure. This means that RAMP can leverage many of the benefits of P2P systems such as dynamic participation (which may encourage more resource owners to devote some or all of their resource to the distributed e-infrastructure when the utilization falls below a certain level). 


Resilience is a key requirement of the RAMP system. Many distributed scheduling systems rely on a single broker component, which results in a single point of failure which can render the whole system unusable. The JADE development environment allows agents to be distributed across a network of machines and incorporates P2P network features to boost system stability and resilience. The key JADE feature utilized by RAMP is the \textit{Main Replication Service}. All JADE agents run within a \textit{container} which provides basic agent communication and management capabilities. JADE requires a \textit{Main container} to act as the control point for the distributed agent system. The Main Replication Service allows the Main container functionality to be replicated amongst a ring of containers, to which normal containers connect. In the RAMP system, each Resource Agent runs in a replicated version of the system's Main container. User Agents run in normal containers; a User Agent's container can connect directly to any Resource Agent's Main container and, via container replication, have access to all Resource Agents in the system. 

In addition, when a new Resource Agent connects to the system, in needs only connect to one of the Main containers to join the whole market place. The JADE \textit{Address Notification Service} runs within all the containers in the system. It monitors agents and therefore containers entering and leaving the system, and reconfigurs the network accordingly. In this way, if one of the Main containers crashes or otherwise exits the system, all of the User Agent containers connected to that Main container are reconfigured to automatically connect to another Main container, and the connections in the Main container ring are suitable adjusted. 

\subsubsection{Agent communication ontology}

The FIPA Agent Communication Language terms discussed in Section \ref{sec:protocol} define the basic semantics of how agents interact in a FIPA compatible multi-agent system. However, these basic actions do not cover the full, rich lexicon which agents need to possess in order to implement the negotiation protocols described earlier. Within FIPA compliant agents, ontologies are used to represent the set of concepts and symbols that agents need to communicate about. This standard method of inter-agent `language' based on a well defined ontology allows agents implemented using different software environments to be mutually intelligible. 

In the RAMP system, agents primarily need to communicate about Requests for Quotations. Therefore we have developed an ontology that allows agents to communicate based around the terms of the RFQL syntax described above. Within JADE, this is realized as a series of Java classes that extend the \texttt{jade.content.onto.Ontology} class, with each class corresponding to different ontological terms. 


\section{System Integration}

AHE and RAMP are independent systems, but are designed to closely interoperate. AHE provides a persistent job launching and execution management service. RAMP provides a market place in which compute cycles can be traded between users and resources. 

The AHE is pre-configured with details of the applications which the user can run, and the static set of resources on which the applications are installed. This is in contrast to the Condor approach, which stages application binaries to resources before they are run. The parallel MPI applications which the AHE was designed to run are often difficult to compile and need an expert user or system administrator to optimize them for a particular machine architecture, which makes binary staging impractical.

\begin{figure*}[!t]
\centering
\includegraphics[scale=0.5]{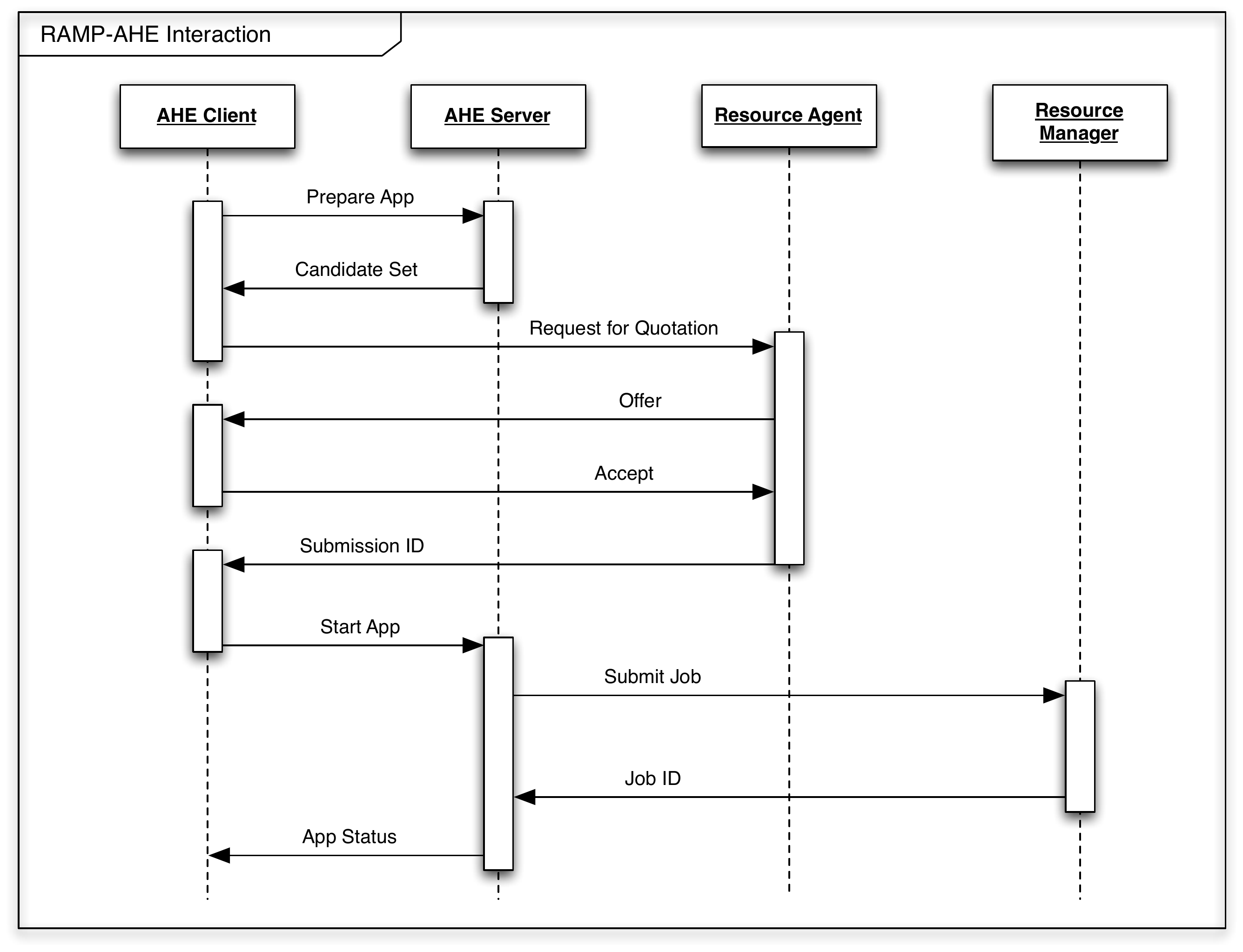}
\caption{The sequence of operations between the AHE client (containing the User Agent), the AHE server, and a single resource, required to launch an application on that resource.}
\label{fig:seq}
\end{figure*}

\section {Evaluating the System}

The RAMP system is designed to manage resource allocation across a set of high performance compute (or cloud) resources. Deploying the system across such a set of resources in order to evaluate the performance and capabilities is impractical, since root access could be required to install tools that interact with the queuing system, machines could be taken down for maintenance periods and so on. Therefore, we found it practical to develop a simulation environment which would allow us to evaluate RAMP without the external difficulties inherent in using a production HPC e-infrastructure. 

\begin{table}[!htbp]
\centering
\begin{tabular}{|c|c|c|c|c|}
 \hline
\textbf{Machine} & \textbf{Cores} & \textbf{Log Start} & \textbf{Log End} & \textbf{Log Duration}\\
  &  & \textbf{Date} & \textbf{Date} & \textbf{(Months)}\\
\hhline{|=|=|=|=|=|}
LLNL Atlas & 9216 & Nov 2006 & Jun 2007 & 8\\
 \hline
LLNL Thunder & 4008 & Jan 2007 & Jun 2007 & 5\\
 \hline
ANL Intrepid& 163840 & Jan 2009 & Sept 2009 & 8\\
 \hline
RICC & 8192 & May 2010 & Sept 2010 & 5\\
 \hline
CEA CURIE & 93312 & Feb 2011 & Oct 2012 & 20\\
\hline
\end{tabular}
\caption{Parallel Workload Archive Project log files used in the simulation environment.}
\label{tab:machineused}
\end{table}

In order to perform a realistic evaluation of the RAMP system, our simulation uses historical resource usage data obtained from a number of high performance computing systems, collected and made available by the Parallel Workload Archive Project\footnote{http://www.cs.huji.ac.il/labs/parallel/workload/}. Archived logs are converted to the Standard Workload Format (SWF) \cite{swf} and in some cases cleaned of erroneous data. 

The logs chosen to base simulations on are summarized in table \ref{tab:machineused}. As can be seen from the table, all of the logs from the archive cover several months of continuous operation. This means that with a relatively few logs we can create a diverse simulated ecosystem with different Resource Agents within the simulated system starting from different time points within the same file. The logs used were chosen to represent a diverse heterogeneous e-infrastructure, with both large petascale machines and smaller clusters.  

The simulation environment consists of a resource plug-in for the Resource Agent which allows it to interact with an historic usage log as if it were a live queuing system on a production HPC resource. This in turn is done through two scripts:

\begin{itemize}
\item \texttt{fake\_qstat.pl}: This script mimics to some extent the behaviour of the qstat command. Qstat, or a differently named variant, is a familiar tool on many HPC systems that allows the user to investigate the status of the queuing system. The script takes as its parameters the path to a configuration file along with the number of cores required by the user and the time the job must run. The configuration file lists the full path to the log file to be read, the time offset where the log should be read from, and also the system start time. This last value allows the simulated system to evolve over time. The system start time represent the beginning of the resource log file. Using the difference between the system start time and the current time (plus the time offset), the script can provide a snapshot of the system state at a given time. This snapshot considers the currently running and queued jobs at the given time to calculate whether sufficient cores will be available for the requested job to start at the time specified by the user. If the job can be run, the script returns the percentage load on that machine at the time the job is to be run.  
\item \texttt{fake\_qrstat.pl}:  This script is almost exactly the same as the \texttt{fake\_qstat.pl} script, but instead of returning a utilization percentage it returns a fake reservation ID if the job can be satisfied, and also inserts the job's details into the machine log. 
\end{itemize}

With these two scripts, our Resource Agent plug-in and our historic queue data, we can investigate various aspects of the RAMP system in a way that closely approximates a real deployment. 

\section{Simulation Environment Setup}
\label{sec:simenv}

\begin{table}[!htbp]
\centering
\begin{tabular}{|c|c|c|c|c|}
\hline
\textbf{System} & \textbf{Base} & \textbf{Time} & \textbf{Start} & \textbf{Min}\\
\textbf{name} & \textbf{system} & \textbf{(sec)} & \textbf{Price} & \textbf{Price}\\
\hhline{|=|=|=|=|=|}
atlas1 & LLNL Atlas & 3370000 & 33 & 25\\
 \hline
atlas2 & LLNL Atlas & 1370000 & 33 & 26\\
 \hline
thunder1 & LLNL Thunder & 250000 & 70 & 40\\
 \hline
thunder2 & LLNL Thunder & 1300000 & 75 & 60\\
 \hline
thunder3 & LLNL Thunder & 130000 & 70 & 35\\
 \hline
thunder4 & LLNL Thunder & 450000 & 75 & 50\\
\hline
intrepid1 & ANL Intrepid & 50000 & 55 & 35\\
 \hline
intrepid2 & ANL Intrepid & 1500000 & 65 & 25\\
 \hline
intrepid3 & ANL Intrepid & 15000000 & 53 & 25\\
 \hline
intrepid4 & ANL Intrepid & 750000 & 55 & 30\\
 \hline
intrepid5 & ANL Intrepid & 2500000 & 65 & 28\\
 \hline
intrepid6 & ANL Intrepid & 90000 & 53 & 30\\
 \hline
ricc1 & RICC & 50000 & 40 & 25\\
 \hline
ricc2 & RICC & 7570000 & 45 & 25\\
 \hline
ricc3 & RICC & 500000 & 45 & 25\\
 \hline
ricc4 & RICC & 757000 & 45 & 30\\
 \hline
curie1 & CEA CURIE & 150000 & 80 & 40\\
 \hline
curie2 & CEA CURIE & 1375000 & 80 & 65\\
 \hline
curie3 & CEA CURIE & 350000 & 80 & 30\\
 \hline
curie4 & CEA CURIE & 2375000 & 70 & 65\\
 \hline
\end{tabular}
\caption{Simulation environment machine setup. The time point is the number of seconds within the log at which the test system started. All systems were started at a minimum of 50000 seconds into the machine log file in order to allow the load on the machine to reach a production level (some machine logs commence with the machine being turned on, meaning that initially the queue is empty).}
\label{tab:exp2setup}
\label{tab:machinesetup}
\end{table}

To evaluate the pure performance of the RAMP system in these initial investigations we deployed all the agents within our simulation environment on a single high powered Ubuntu Linux workstation, to eliminate performance problems that could be introduced by network bandwidth limitations between machines. We took initial configurations based on the machine log files listed in table \ref{tab:machineused}, and working from different time points within those logs, configured an experimental system made up of twenty resources. 

The configuration of those resources is shown in table \ref{tab:machinesetup}, with a single Resource Agent within the system representing each system.  The Resource Agents within the simulation environment logged all transactions for further analysis.  The system was left to evolve in real time as the experiments were performed. 

The static resource parameters for each Resource Agent were all assigned the same default configuration. We did this so that decisions made in the investigations we performed were governed by machine load and price, rather than static constraints. 

\section{Using the RAMP System}
\label{sec:inv1}

The first tests we performed were designed to assess the RAMP system's ability to successfully schedule a range of heterogeneous workloads. Specifically, we investigated the following two aspects of the RAMP system:

\begin{itemize}
\item \textbf{Investigation 1}: Initially we want to confirm that the RAMP system works, that jobs can be placed with resources at prices favourable to both the user and the resource, and that a majority of jobs submitted will be successful. 
\item \textbf{Investigation 2}: Secondly, we wish to assess how efficiently jobs are placed within the system. 
\end{itemize}

In order to perform these investigations, we submitted a range of different jobs, in terms of system requirements, core counts and deadlines, which are listed in table \ref{tab:experimentsetup}, to our RAMP simulation environment. This range of jobs represents the typical tasks a HPC machine will be put to, from small scale, short running jobs to large, long running capability workloads. 

\begin{table}[!htbp]
\centering
\begin{tabular}{|c|c|c|c|}
 \hline
\textbf{Experiment} & \textbf{Cores} & \textbf{Start time} & \textbf{Price}\\
\hhline{|=|=|=|=|}
exp1 & 16 & 5 min & 70\\
 \hline
exp2 & 16 & 60 min & 55\\
 \hline
exp3 & 16 & 12 hours & 35\\
 \hline
exp4 & 256 & 5 mins & 50\\
 \hline
exp5 & 256 & 60 min & 30\\
 \hline
exp6 & 256 & 12 hours & 25\\
 \hline
exp7 & 1024 & 60 min & 55\\
 \hline
exp8 & 1024 & 12 hours & 35\\
 \hline
exp9 & 4096 & 60 min & 55\\
 \hline
exp10 & 4096 & 12 hours & 35\\
 \hline
 exp11 & 20480 & 5 min & 80\\
 \hline
exp12 & 20480 & 60 min & 55\\
 \hline
exp13 & 20480 & 12 hours & 35\\
 \hline
\end{tabular}
\caption{Details of experimental workloads run on the RAMP simulation environment.}
\label{tab:experimentsetup}
\end{table}

Each job listed in table \ref{tab:experimentsetup} was run consecutively, using a separate instance of the User Agent. Each run of all thirteen jobs was repeated three times to better control for temporal anomalies within the system. In each case the User Agent was configured to conduct three rounds of bidding. 

\subsection{Results}
\label{sec:exp1res}

\begin{figure*}[t]
    \centering
    \begin{subfigure}[b]{0.5\textwidth}
        \centering
	\includegraphics[width=\textwidth]{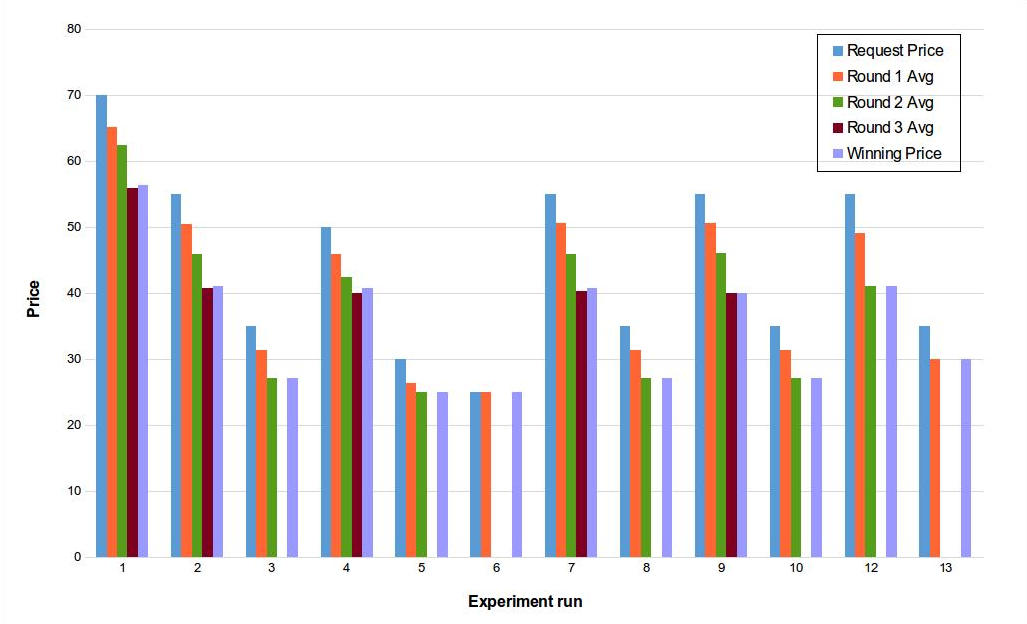}
        \caption{}
\label{fig:exp1-price}
    \end{subfigure}%
        ~ 
      \begin{subfigure}[b]{0.5\textwidth}
        \centering
	\includegraphics[width=\textwidth]{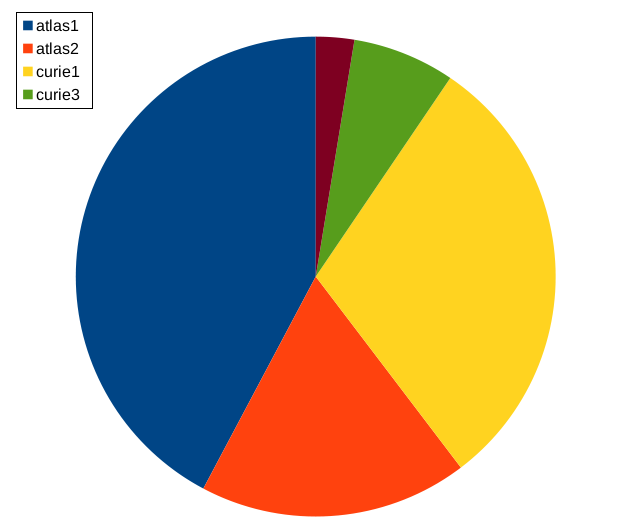}
        \caption{}
\label{fig:exp1-alloc}
    \end{subfigure}%
    \caption{(a) Request, offer and winning prices for the requests shown in table \ref{tab:experimentsetup} for three rounds of bidding. Where bidding round values are not shown, no offers were made in that round. Experiment run 11, which failed, is not shown; (b) Resource \textit{attractiveness} plotted over time. Attractiveness is analogous to the value by which a resource is willing to reduce its prices.}
\end{figure*}

From the resulting logs generated by the User Agents and Resource Agents, we computed mean prices for each of the bidding rounds and also the winning price. 
Figure \ref{fig:exp1-price} displays these results as a bar chart, showing how the price offered by each resource fell with each round of bidding. 


Where bidding rounds show no result, either the price of the job had fallen below the minimum price threshold of a resource in the system, or the system load on has increased on the resource, meaning that no resource is willing to continue bidding. We found that the median offer price was 71.1\% of the request price.  

In order to investigate the efficiency of our system mapping jobs onto resources, we also examined which of the resources in the system won each of the auctions outlined in table \ref{tab:experimentsetup}. Five of the resources won all of the jobs submitted in three experimental repetitions, with the share of the jobs distributed as shown in figure \ref{fig:exp1-alloc}.

\begin{figure*}[t]
    \centering
    \begin{subfigure}[b]{0.47\textwidth}
        \centering
	\includegraphics[width=\textwidth]{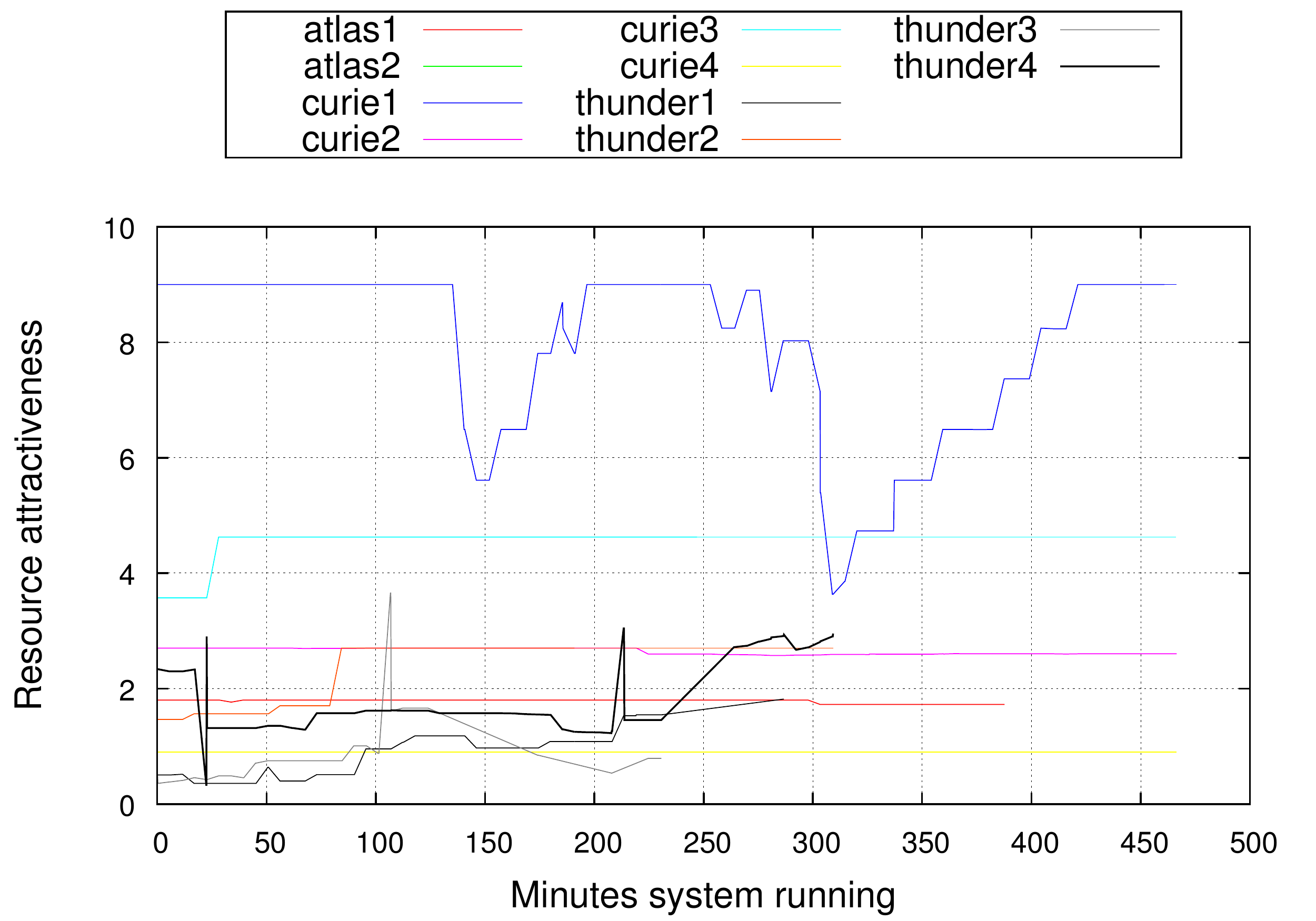}
        \caption{}
        \label{fig:exp1-attract}
    \end{subfigure}
~ 
     \begin{subfigure}[b]{0.47\textwidth}
        \centering
	\includegraphics[width=\textwidth]{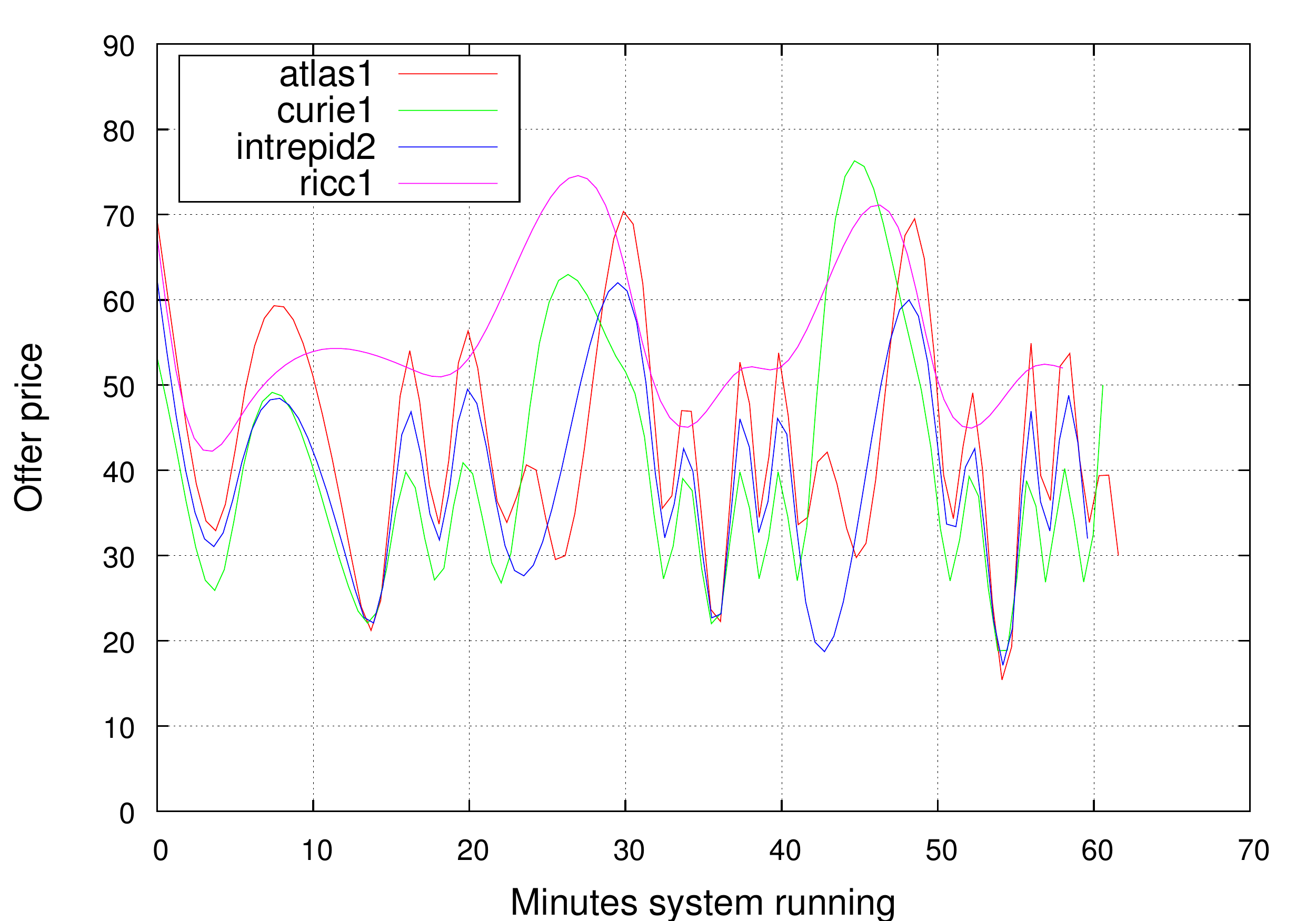}
        \caption{}
        \label{fig:exp1-pricevariation}
    \end{subfigure}
    \caption{(a) The winning resources for the auctions of workloads listed in table \ref{tab:experimentsetup}. Five resources were allocated all of the jobs submitted; (b) The change in resource offer prices over time for four resources in the system.}
\end{figure*}

To assess whether these resources were allocated to the most appropriate resource, and thus the overall efficiency of the system, we need to consider just more than just the load on each system. Allocation of workloads to resources within the competitive RAMP market place is based on a function combining both the load on the machine and the price it is willing to offer to get jobs, as outlined in Section \ref{sec:primod}. 

Therefore, to assess whether work is allocated to resources efficiently, we need to consider the \textit{attractiveness} of the resource to the users of the system. A useful measure of the attractiveness of the resource is the price by which it is willing to reduce its offer price while bidding, since this is based on the start and minimum price configuration of the machine, and its load. 

In figure \ref{fig:exp1-attract} we plot the attractiveness of the resources in the system over the evolution of the simulation environment. Resources not included in the plot were loaded to such an extent that they did not make offers. 

\subsection{Discussion of Results}

The results plotted in figure \ref{fig:exp1-price} confirm that our system is capable of successfully allocating jobs to resources at prices lower than the user is willing to pay. Although the majority of requests were allocated during our experimental runs, it is likely that this will not always be the case, especially where the user sets their opening price below the minimum price of all resources in the system, or where resource load across the platform is sufficient that large jobs that need to start very soon cannot be accommodated. The failure of experiment run 13 shows us that very large requests that need to be run soon will likely fail, even when the user is willing to pay a premium to execute the workload. 

We found that resource request prices that were very close to the minimum system prices offered by the majority of resources resulted in fewer rounds of bidding, but achieved better prices for the user. 

In a production environment, a user would not necessarily know the minimum prices set by the resource providers but, with experience, may well come to learn reasonable estimates of the minimum price various resources were willing to offer, and would likely be able to price their workloads accordingly. 

Our assessment of the efficiency of the system shows that the jobs are allocated to the most attractive resources. As shown in figure \ref{fig:exp1-alloc}, all of the test workload requests submitted to the system were allocated to just five of the available twenty resources (with the exception of the request that failed). Looking at figure \ref{fig:exp1-attract} we see that in general the most attractive resources won the resource requests made. 

Though the curie1 and curie3 resources remained the most attractive over the simulation execution duration, they did not take all of the jobs, and were not even the biggest winner, which was the atlas1 machine. This can be explained by the fact that temporal changes within the attractiveness of resources mean that at different points, one resource will be more attractive than others. Also, the User Agent accepts offers on a first come first serve basis, so that if two resources make the same offer, the offer received first by the Resource Agent will be favoured. This means that a single resource with a low load and favourable pricing structure does not completely dominate the platform and take all requests made. 

\section{Investigating Job Pricing}

The Resource Agent provides a bespoke spot price for the resource it is managing, in response to requests from User Agents. As such, the price offered by a resource fluctuates over time. Figure \ref{fig:exp1-pricevariation} shows how the price for various resources varies over time, while the simulation environment is running.

The prices agreed for resource requests are governed by a combination of parameters under the control of both the Resource Agent and the User Agent. The Resource Agent is configured with a minimum price and a start price, which are used, along with the load on the machine, to generate a bid reduction price. Choosing start and minimum prices is therefore a a task which the machine administrators need to invest some effort in. 

The key is to set start and minimum prices that will actually result in bids that, on average, meet the cost that the resource owners want to sell their cycles for. Of course, the minimum price provides a lower bound; when auctions fall below this minimum, the resource will refuse to participate further in the auction. 

In addition to the configuration parameters set for the Resource Agent, the price is also governed by how many rounds of bidding the User Agent specifies. To optimize these parameters we performed the following investigations:

\begin{itemize}
\item \textbf{Investigation 1}: We calculated the average offer price made by a resource over a period of system operation, and compared it to the start and minimum price used to configure the resource. 
\item \textbf{Investigation 2}: We performed an experiment to discover the optimum number of bidding rounds required to minimize the price paid by the user. 
\end{itemize}

The resource offer price is also governed by the price that other resources offer in the system (which is outside the control of any given resource administrator) and the prices that users of the system are willing to pay for the workloads they need to execute. 

\begin{figure}
\centering
\includegraphics[scale=0.2]{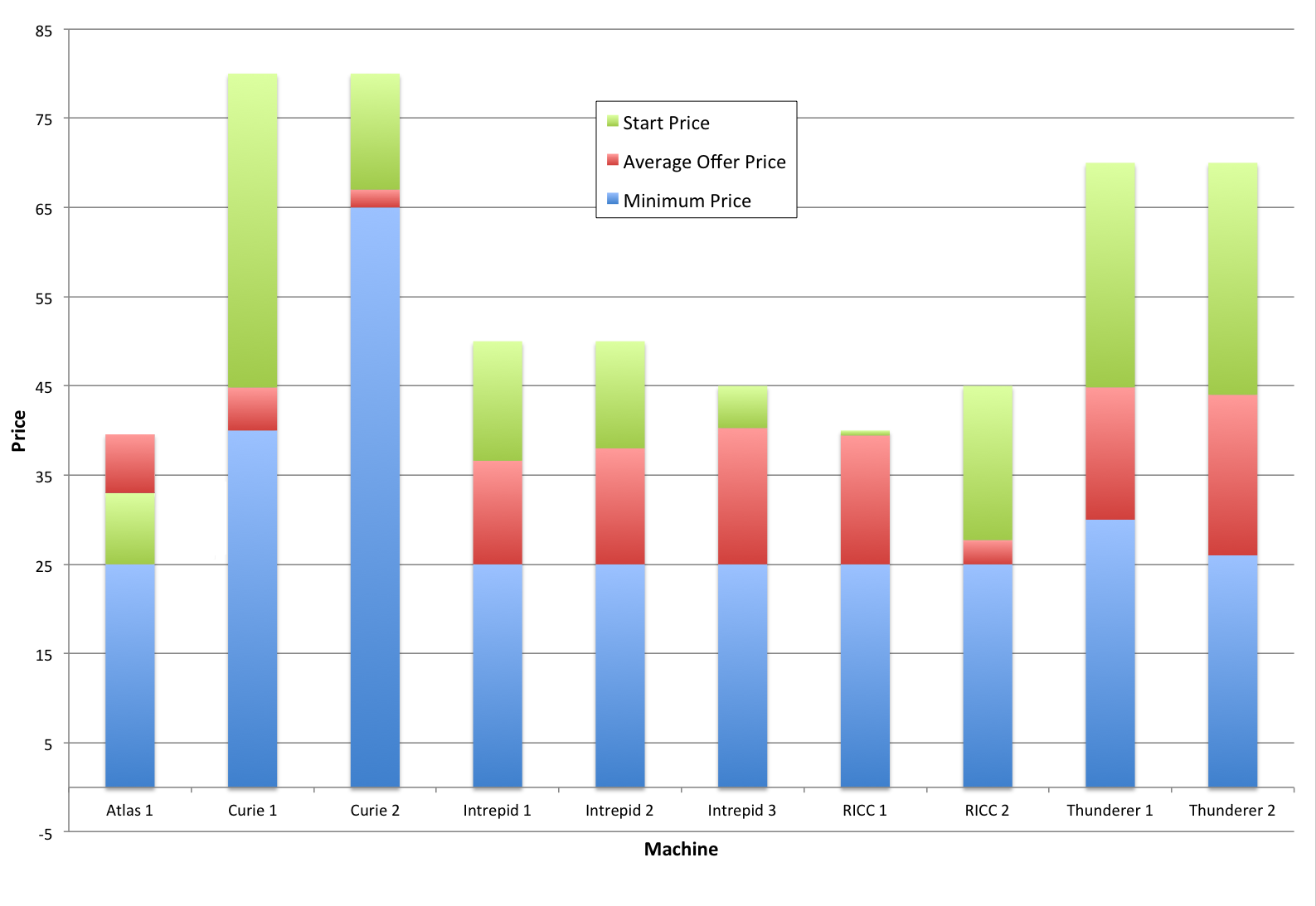}
\caption{Comparison of the average offer price made by each resource to its minimum price and start price.}
\label{fig:exp1-pricerange}
\end{figure}

We analysed the results for the workloads presented in Section \ref{sec:inv1} to discover the mean offer price made by each resource in the system. In figure \ref{fig:exp1-pricerange} we plot this mean price alongside the start and minimum prices. 

To determine the optimum number of bidding rounds required to minimize the price paid by the user, we use the environment simulation described in Section \ref{sec:simenv} to make a user requisition for a single auction unit. The request was run ten times, with each run executed with an increasing number of auction rounds, from one round to ten rounds. We repeated each experiment run three times, and calculated mean values for the sale price and duration of the auction, shown in table \ref{tab:exp3results}.

\begin{figure*}[t]
    \centering
    \begin{subfigure}[b]{0.5\textwidth}
        \centering
	\includegraphics[width=\textwidth]{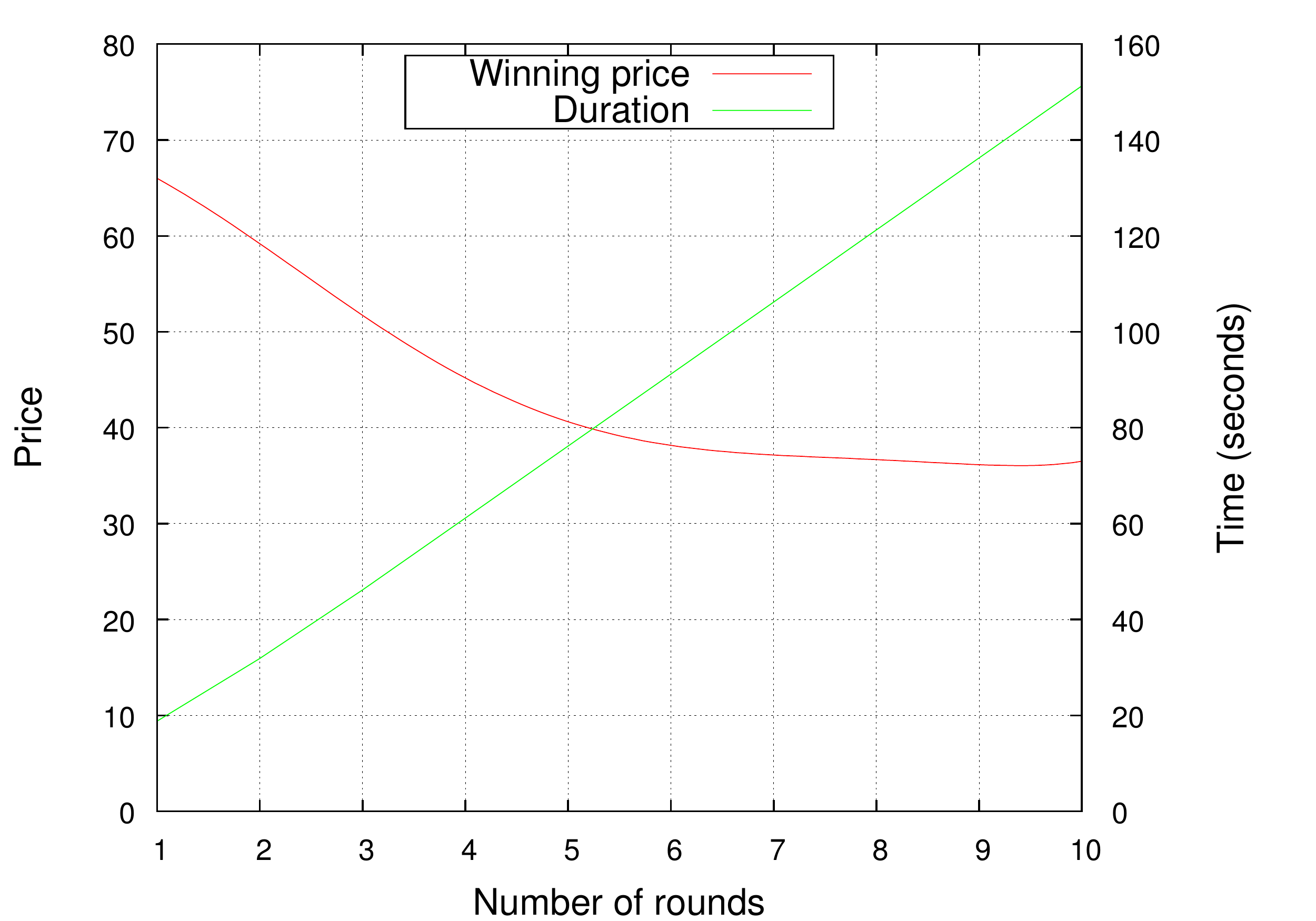}
        \caption{}
	\label{fig:exp3-1}
    \end{subfigure}%
        ~ 
    \begin{subfigure}[b]{0.5\textwidth}
        \centering
	\includegraphics[width=\textwidth]{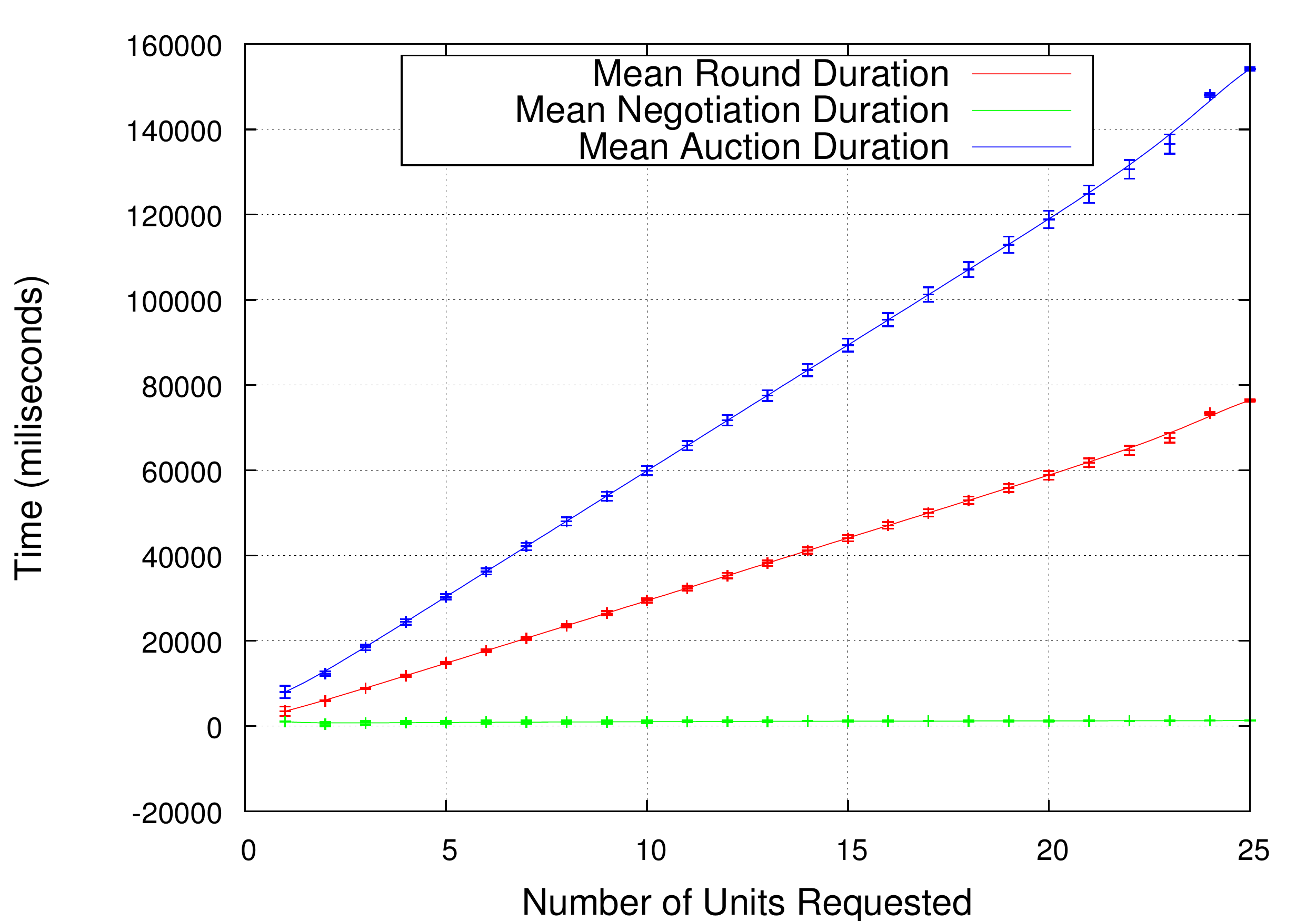}
        \caption{}
\label{fig:exp2}
    \end{subfigure}
    \caption{(a) Plot showing how the increase in number of bidding rounds affects final cost price and auction completion time; (b) Plot showing how the increase in units per auction affect system performance. As the number of units increase, the time taken to complete the auction scales linearly. Bars show the standard deviation.}
\end{figure*}

In figure \ref{fig:exp3-1} we plot the mean sale price and the mean auction duration the for auctions with one to ten rounds of bidding. 

\begin{table}[t]
\centering
\begin{tabular}{|c|c|c|}
\hline
\textbf{Number of Rounds} & \textbf{Mean Sale Price} & \textbf{Mean Duration (sec)}\\
\hhline{|=|=|=|}
1 & 66.00 & 18.81 \\
\hline
2 & 60.00 & 31.91 \\
\hline
3 & 51.83 & 46.19 \\
\hline
4 & 41.83 & 61.15 \\
\hline
5 & 37.40 & 76.18 \\
\hline
6 & 35.40 & 91.18 \\
\hline
7 & 37.83 & 106.21 \\
\hline
8 & 37.83 & 121.24 \\
\hline
9 & 34.75 & 136.26 \\
\hline
10 & 36.50 & 151.26 \\
\hline
\end{tabular}
\caption{Offer price and auction duration as the number of bidding rounds increases.}
\label{tab:exp3results}
\end{table}

\subsection{Discussion of Results}

Selecting optimal configuration parameters for a resource is a complex task. As figure \ref{fig:exp1-pricerange} shows, offers made by a resource will roughly fall between the start and minimum prices set of the resource. However, the minimum price is seemingly the most important parameter; where the mean price is set at a level comparable to other resources in the system, the mean offer price will usually be comparable to those other resources too, and somewhat higher than the minimum price.  However, setting a high starting price will increase the attractiveness of the resource by increasing the amount which the resource is willing to reduce its offers by while bidding. In summary, to avoid being outbid and to increase the chance of auction success, a resource owner should try to set a minimum price around the same level to other resources in the system, but a high starting price. 

As we see from figure \ref{fig:exp3-1} and table \ref{tab:exp3results}, as the number of auction rounds used by the User Agent increases, the final offer price accepted is reduced, with a tail off at five rounds, suggesting that users wanting to optimize the price they pay for auction units could do so by running auctions with five bidding rounds. However, as the number of auction rounds increases, so does the time taken to complete the auction. This scales linearly with the number of auction units, as is to be expected since auction rounds are of a fixed duration. Users must take into account this trade-off when initiating multi-round auctions. 


\section{Investigating RAMP Performance}

Our multi-unit auction system is, we believe, unique, but it is also important that it is usable. A key aspect of usability is the responsiveness of the system. It is important that the RAMP system responds well to user requests, and scales both with the number of units an individual is requesting (in a multi-unit auction) and with the number of simultaneous users of the system. To assess this performance, we conducted two investigations into system scalability: 

\begin{itemize}
\item \textbf{Investigation 1}: We measured the performance of the system in terms of auction duration as the number of request units within a combinatorial reverse auction increase. 
\item \textbf{Investigation 2}: We measured the performance of the system in terms of auction duration and average system response time as the number of User Agents participating in the system increases. 
\item \textbf{Investigation 3}: To assess the impact of network performance on a the responsiveness of a real-world deployment of RAMP, we repeated investigation 2 with our Resource Agents deployed across a network of machines. 
\end{itemize}

\subsection{Results}

Using the simulation environment outlined in table \ref{tab:exp2setup} we sequentially submitted requests via a single Resource Agent in order to investigate how performance increases with the number of auction units. With each submission the number of units within the request increased, meaning the terms within the combinatorial reverse auction increased.  

The simulation environment was run on a single workstation to eliminate disruptions caused by network problems, with a separate Resource Agent for each simulated resource. Runs were performed for auctions with 1 to 25 units (the deadline and price of each unit was randomly generated), and the whole set of runs was repeated one hundred times, and mean response times calculated. 

The user configurable auction round duration parameter was set to five minutes, so that we could measure round duration without the auction ending. The times taken to complete bidding rounds, negotiate the final auction agreement, and the total time taken for the auction to complete are displayed in figure \ref{fig:exp2}.


To assess how the system performs when multiple individual users are using it, we performed an experiment using the simulation environment outlined above, whereby we ran multiple User Agents simultaneously, each making a single unit request. We ran from 1 to 30 User Agents consecutively, and repeated each run three times then calculated mean response times. Again, the User Agents were configured with a maximum bidding round duration of five minutes, so that auction rounds would not time out before all Resource Agents had been able to respond. 

\begin{figure*}[t]
    \centering
    \begin{subfigure}[b]{0.5\textwidth}
        \centering
	\includegraphics[width=\textwidth]{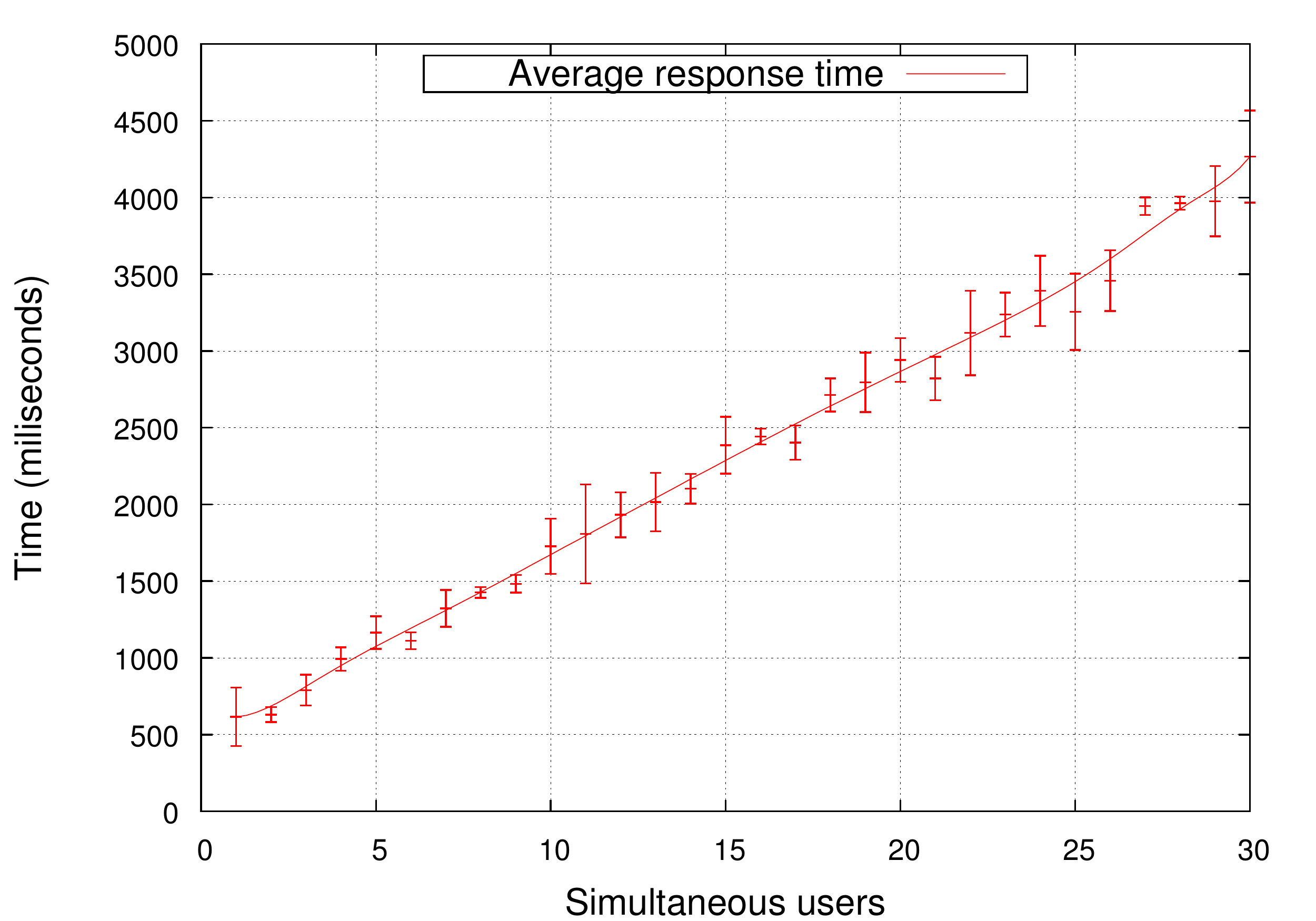}
        \caption{}
	\label{fig:exp4-1}
    \end{subfigure}%
        ~ 
    \begin{subfigure}[b]{0.5\textwidth}
        \centering
	\includegraphics[width=\textwidth]{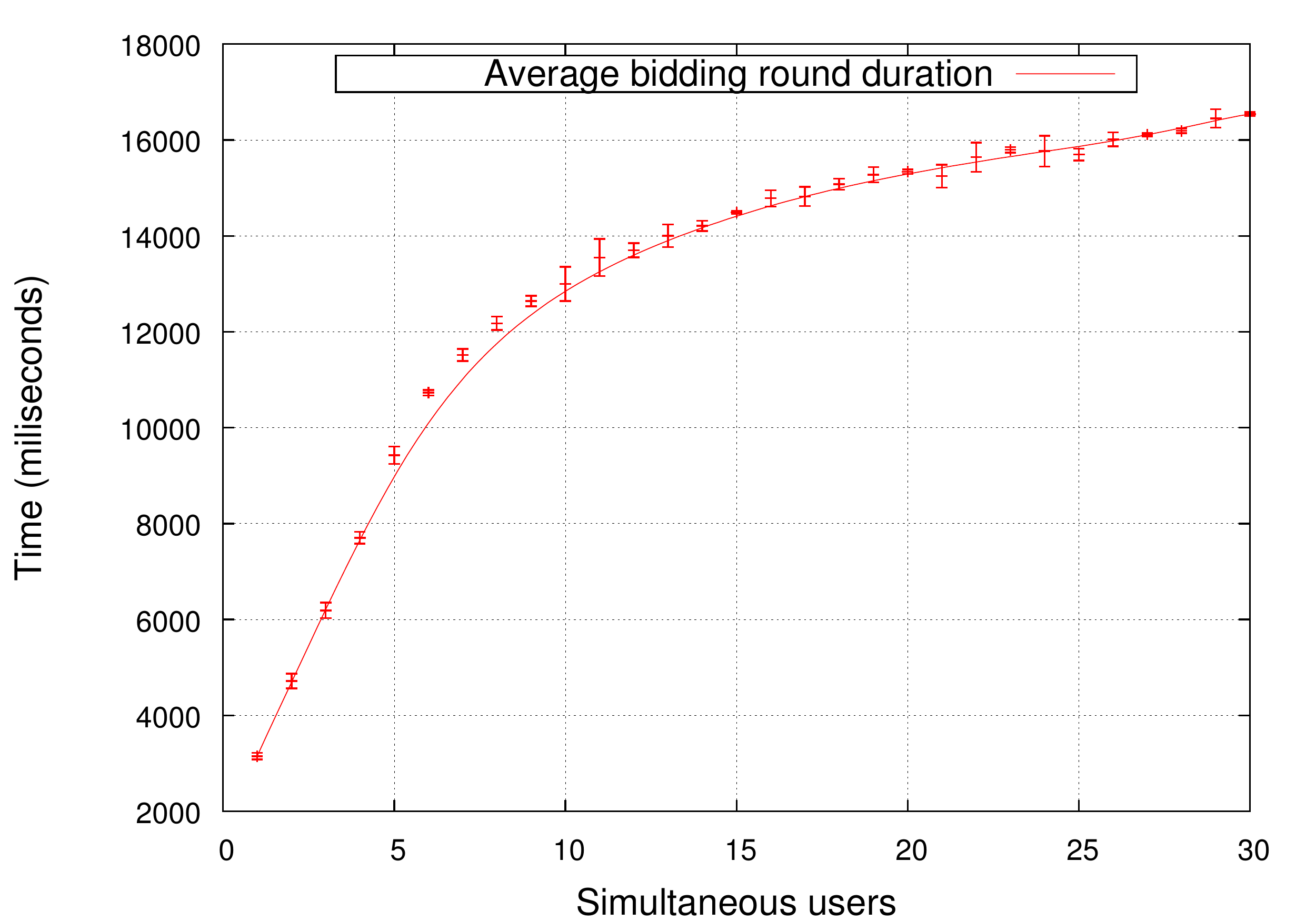}
        \caption{}
	\label{fig:exp4-2}
    \end{subfigure}
    \caption{(a) Plot showing how the increase in competing agents affects mean system response time. The response time scales linearly with the number of concurrent User Agents. Bars show the standard deviation; (b) Plot showing how the increase in competing agents affects mean auction round duration. The round duration rises sharply up to 10 simultaneous users, then tails off. Bars show the standard deviation.}
\end{figure*}

We measured the mean time taken for a Resource Agent to respond to an individual request (shown in figure \ref{fig:exp4-1}) and how the number of competing agents within the system affects the duration of an auction (shown in figure \ref{fig:exp4-2}). 


In real world applications, the RAMP system is intended to be deployed across a network of HPC class resources. Our performance tests so far have only measured performance with RAMP deployed using our simulation environment on a single machine. To ensure that network effects will not adversely affect the performance of the system, we repeated our tests on network deployment of RAMP. 

The system was deployed across 15 networked servers. In a real world deployment we expect that RAMP would be deployed across an Internet wide set of HPC resources. The impracticalities of securing access to such resources in order to carry out our performance tests led to us deploying a system with 10 servers located within the Centre for Computational Science research lab in University College London, with additional Resource Agents deployed at CINECA (Italy), Cyfronet (two agents) and PSNC (both Poland), University of Sheffield (UK). The Resource Agents used the first 15 resource configurations listed in table \ref{tab:exp2setup}.

\begin{figure*}[t]
    \centering
    \begin{subfigure}[b]{0.5\textwidth}
        \centering
	\includegraphics[width=\textwidth]{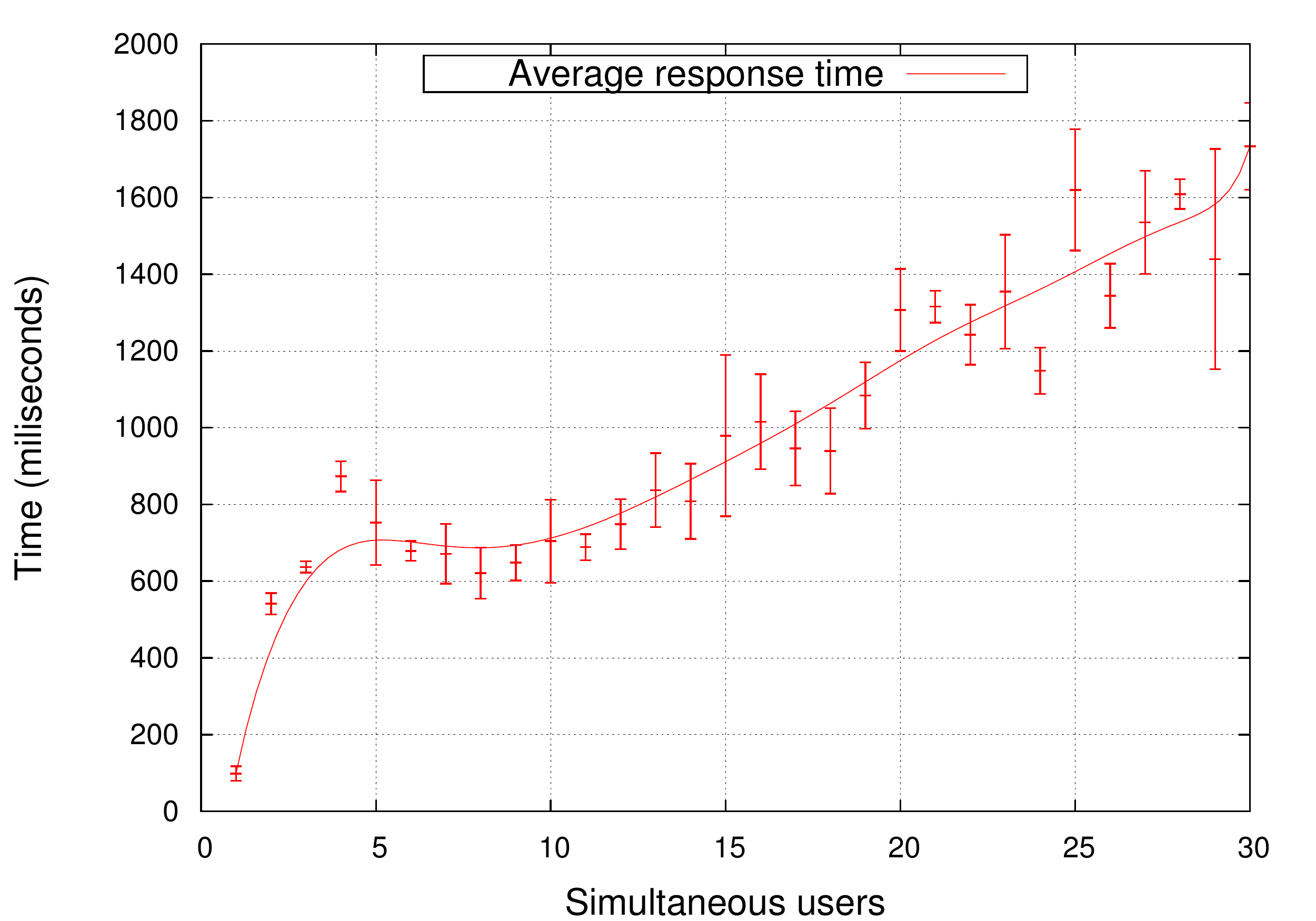}
        \caption{}
        \label{fig:exp7-1}
    \end{subfigure}%
        ~ 
    \begin{subfigure}[b]{0.5\textwidth}
        \centering
	\includegraphics[width=\textwidth]{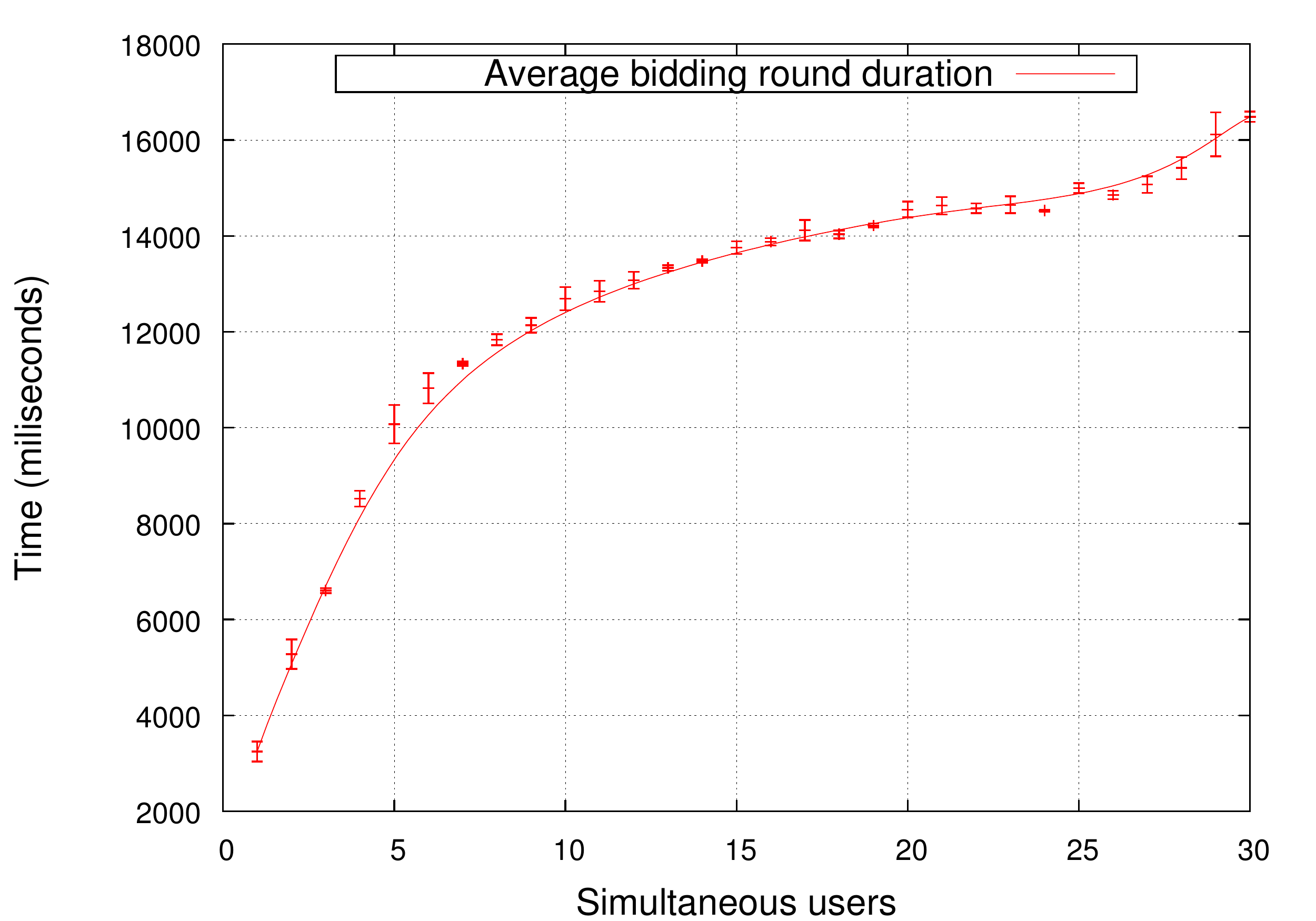}
        \caption{}
        \label{fig:exp7-2}
    \end{subfigure}
    \caption{(a) Plot showing how the increase in competing agents affects mean system response time with RAMP system deployed across a network of resources. Bars show the standard deviation; (b) Plot showing how the increase in competing agents affects mean auction round duration with RAMP system deployed across a network of resources. Bars show the standard deviation.}
\end{figure*}

We repeated the previous investigation, running between 1 and 30 User Agents simultaneously and measuring the impact of doing so on the mean time taken for a Resource Agent to respond to an individual request (shown in figure \ref{fig:exp7-1}) and how the number of competing agents affects the duration of an auction (shown in figure \ref{fig:exp7-2}).


\subsection{Discussion of Results}

As we see from figure \ref{fig:exp2}, the duration of the bidding rounds in an auction scales linearly with the number of units in an auction. This is to be expected, since the duration of the auction is increased by the number of requests the User Agent has to make. Surprisingly, the time taken to negotiate the auction does not increase with the number of units. Adding an additional unit only adds a couple of seconds to the overall duration of the auction so this is unlikely to be of too much concern to the user. 



As the number of simultaneous users using the system increases, the responsiveness of the Resource Agents scales linearly, as we see from figure \ref{fig:exp4-1}. However, as shown in figure \ref{fig:exp4-2}, the time taken to complete an auction increases steeply with the first ten simultaneous users of the system, and then tails off as the number of users increases. The tailing off is well below the maximum auction duration we configured in the system, so we are not seeing the effect of this parameter. It is unclear why figure \ref{fig:exp4-2} is so shaped, and further work is required to understand the observations presented here.  On the whole, the results we have obtained indicate that our system shows good responsiveness and scalability as both the number of auction units and number of users increase. When repeated using Resource Agents deployed across a network of hosts (approximating a real-world deployment of RAMP) we found that the effects of the network did not have a significant impact on performance and similar scaling characteristics were obtained, although there was greater variance in the results obtained. 



\section{Conclusions}

We believe that distribute e-infrastructure platforms (clouds and grids) currently suffer from usability issues that prevent them from being exploited in a systematic fashion. To help improve usability, we have developed AHE, a tool to allow users to focus on applications rather than machines. However, this does nothing to improve the users' total time to solution, and hence needs to be coupled with a resource allocation mechanism. 


In this paper we have presented our investigations into the performance and capabilities of our RAMP resource allocation platform. We have shown that the system is capable of successfully allocating workloads to computational resources, optimising the price that the user pays and selecting the most attractive resources from the set of available machines. The decentralized nature of the system means that it does this without incurring the overheads and failure points present in a centralized brokering system where a single component is responsible for allocating jobs throughout a distributed e-infrastructure. 

Our system shows good performance and scalability, even when deployed across a wides area network of machines. The ability of users to control the number of and duration of bidding rounds means that they can minimize the price they pay while at the same time placing an upper bound on the time taken to achieve a result. The next step we plan to take is to evaluate our RAMP system formally, by conducting a usability study using real users, on a deployment of RAMP across a production e-infrastructure. 

\section{Acknowledgements}

The work described herein has been funded by EPSRC under the RealityGrid project (GR/R67699), the EU FP7 \textit{VPH-Share} (no 269978), \textit{VPH-NoE} (no 223920), \textit{MAPPER} (no 261507) and \textit{ComPat} (no 223979) projects, the EPSRC \textit{Rapid Prototyping of Usable Grid Middleware} (GR/T27488/01) grant, and also by OMII under the Managed Programme \textit{Robust Application Hosting in WSRF::Lite (RAHWL)} project.






%
%
%

\end{document}